\documentclass[aps,10pt,reprint,groupedaddress,superscriptaddress, prx, nofootinbib]{revtex4-2}

\usepackage{amssymb}
\usepackage{graphicx}
\usepackage{amsmath}
\usepackage{multirow}
\usepackage{blkarray}
\usepackage{nicematrix}
\usepackage{tikz}
\usepackage{mathrsfs}
\NiceMatrixOptions{
code-for-first-row = \color{black} ,
code-for-last-row = \color{black} ,
code-for-first-col = \color{black} ,
code-for-last-col = \color{black}
}

\usepackage{braket}
\usepackage{babel}
\usepackage{soul}
\usepackage[compat=1.1.0]{tikz-feynman}
\usepackage[normalem]{ulem}
\usepackage{physics}

\usepackage{mathtools}

\definecolor{mypurp}{rgb}{0.35, 0, 0.7}

\usepackage{amsthm}
\usepackage{txfonts}

\theoremstyle{definition}

\newcommand{\tTC}{\text{TC}}

\usepackage{hyperref}
\hypersetup{colorlinks,breaklinks,
            urlcolor=[rgb]{0.25,0.41,0.88},
            linkcolor=[rgb]{0.25,0.41,0.88},
            citecolor=[rgb]{0.25,0.41,0.88}}

\begin{document}

\def\papertitle{{Information-theoretic principle of emergent 1-form symmetries}}

\title{\papertitle}

\newcommand{\TUM}{\affiliation{Technical University of Munich, TUM School of Natural Sciences, Physics Department, 85748 Garching, Germany}}
\newcommand{\MCQST}{\affiliation{Munich Center for Quantum Science and Technology (MCQST), Schellingstr. 4, 80799 M{\"u}nchen, Germany}}
\newcommand{\MIT}{\affiliation{Center for Theoretical Physics, Massachusetts Institute of Technology, Cambridge, MA 02139, USA}}

\author{Yu-Jie Liu} 
\thanks{These authors contributed equally.\\}
\TUM \MCQST \MIT
\author{Wen-Tao Xu} 
\thanks{These authors contributed equally.\\}
\TUM \MCQST
\author{Frank Pollmann}  \TUM \MCQST
\author{Michael Knap}  \TUM \MCQST

\makeatletter
\renewcommand{\@makefnmark}{\textsuperscript{\arabic{footnote}}}
\renewcommand{\thefootnote}{\arabic{footnote}}
\makeatother

\begin{abstract}
    Higher-form symmetries act on sub-dimensional spatial manifolds of a quantum system. They can emerge as an exact symmetry at low energies even when they are explicitly broken at the microscopic level, making them difficult to characterize. In this work, we propose that the emergence of 1-form symmetries is information-theoretic in nature, characterized by the preservation of information about a specific bare (microscopic) 1-form symmetry. As a consequence, the loss of the emergent 1-form symmetry is an information-theoretic transition which we argue to be revealed from the long-range entanglement in the ensemble of post-measurement states. We analytically determine the regimes in which a 1-form symmetry emerges in product states on one- and two-dimensional lattices. 
    In analytically intractable regimes, we demonstrate how to efficiently detect 1-form symmetries with a global quantum error correction (QEC) decoder and numerically examine the information-theoretic transition of the 1-form symmetry, including systems with $\mathbb{Z}_2$ topological order. As a practical application of our framework, we show that once the 1-form symmetry is detected to exist, a topological quantum phase transition  characterized by the spontaneous breaking of the 1-form symmetry can be  accurately determined by a disorder parameter. We further argue that our proposed theory for emergent 1-form symmetries offers new perspectives on particle condensation and suggests sharp information-theoretic phase boundaries between Higgs and confining regimes in the $\mathbb{Z}_2$ lattice gauge theory.    
    %By exploiting ideas from quantum error correction, our work develops an information-theoretic criterion for emergent 1-form symmetries, which furthers our understanding of exotic symmetries and offers practical routes toward their characterization. 
    
\end{abstract}

\maketitle

\tableofcontents

\section{Introduction}

Symmetries are fundamental for identifying and classifying different phases of matter. Generalizations of symmetries beyond the usual global on-site symmetries have revealed surprising consequences rooted in previously unappreciated structures of physical systems~\cite{High_form_Kapistin_2015,McGreevy_2023,Bhardwaj:2023,sal:2024a}. Among these are higher-form symmetries. In $d$ dimensions, a $p$-form symmetry has symmetry operators supported in $d-p$ dimensional manifolds and charged operators supported in $p$ dimensions. The familiar example of a conventional global on-site symmetry is in this classification a 0-form symmetry, which acts on the whole system and has local charged operators. 
In contrast to 0-form symmetries, higher-form symmetries ($p\geq1$) then act on subdimensional manifolds. A paradigmatic example are the 1-form symmetries given by Wilson and 't Hooft loop operators in a $\mathbb{Z}_2$ gauge theory in $d=2$. While conventional 0-form symmetries are exact properties of the system, determined by an operator that commutes with the Hamiltonian, higher-form
symmetries posess an enhanced robustness; even
when the bare symmetry is absent at the microscopic
level, it can still exist as an exact emergent symmetry at low
energies~\cite{Hastings_and_Wen_2005,Self_dual_critically_Ising,Wen_emergent_high_form_2023, cherman:2024,sal:2024}. In other words, higher-form symmetries remain robust even under perturbations that explicitly break the bare symmetry, and their exact expression in general depends on the specific details of those perturbations. This aspect challenges the direct characterization of higher-form symmetries in physical systems.  Despite recent progress, a precise definition for emergent higher-form symmetries in a quantum state and a general understanding of when they cease to exist under large perturbations remain elusive.

Beyond their theoretical significance, higher-form symmetries also offer practical insights into the characterization of quantum many-body systems. They are known to play a fundamental role in exotic phases of matter, in particular phases with topological order~\cite{Zohar_2004,Hastings_and_Wen_2005,NUSSINOV_2009, FQHE_1982,Laughlin_1983,kitaev_2002,String_net_2005,Kitaev_2006,Zoo_2017}. The emergence of topological order can be understood as either spontaneous breaking of higher-form symmetries or a mixed anomaly between them~\cite{High_form_Kapistin_2015,High_form_wen_2019,Wen_emergent_high_form_2023,McGreevy_2023}. One standard method for detecting topological order is to measure topological entanglement entropy~\cite{TEE_Kitaev_Preskill_2006,TEE_levin_Wen_2006, LRE_2010}, but this is experimentally difficult because it requires a number of measurements that grows exponentially with the perimeter of the subsystem of interest~\cite{TC_quantum_computer_2021}. Alternatives include measuring the Fredenhagen-Marcu (FM) string order parameter~\cite{FM_1983,FM_1986} or the perimeter law for Wilson or ’t Hooft loops~\cite{Wegner_duality_1971,Kogut_1979}, though these also demand exponentially large sample sizes and rely on the presence of higher-form symmetries~\cite{xu_2024_FM}. As synthetic quantum systems are becoming increasingly capable of realizing exotic quantum states with abelian~\cite{TC_quantum_computer_2021,Cochran2024,TC_Quantum_simulator_2021}, non-abelian~\cite{D4_trapped_ion_2024, doublefib:2024, Evered2025}, and even non-equilibrium~\cite{Will2025} topological order, there is a need for more efficient detection protocols. This raises the question of whether there exists an efficient, more direct framework to utilize higher-form symmetries for detecting topological order in practice.

\begin{figure}
    \centering
    \includegraphics[scale = 1]{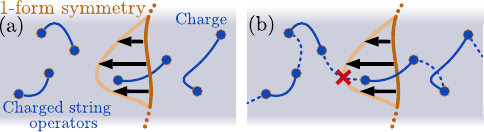}
    \caption{\textbf{A geometric picture for emergent 1-form symmetries.}  We decompose the quantum state as a 1-form symmetric state acted on by charged string operators.\protect\footnotemark[\value{footnote}] (a) When the choice of the charged string operators (blue) for each charge configuration is unique, %(up to smooth deformation in two and higher dimensions, see Sec.~\ref{sec:2dproduct} and Sec.~\ref{sec:1form_qec})
    we can deform the 1-form symmetry (brown), such that it avoids the intersection with the string operator; the 1-form symmetry then effectively commutes with the charged string operators. (b) Assuming there exists no unique choice of the charged string operators for the same charge configuration (solid and dashed lines), the deformation will necessarily intersect with one set of the strings, preventing the deformed 1-form symmetry operators from effectively commuting with them. Therefore, when both configurations appear with comparable probability, one cannot commute the 1-form symmetry through the charged strings and the 1-form symmetry is absent. %For better visualization, we sketch the scenario for a 2D system.
    }
    \label{fig:pulling}
\end{figure}

In this work, we focus on 1-form symmetries ($p=1$) and argue that the existence of an emergent 1-form symmetry is intrinsically tied to the decodability of the string operators creating the 1-form symmetry charges. The decodability determines whether the 1-form symmetry effectively commutes with these charged string operators (see Fig.~\ref{fig:pulling}). Given a fixed reference representation of a bare 1-form symmetry acting on the microscopic Hilbert space and exactly respected by the unperturbed Hamiltonian, our framework asks whether the perturbed state retains information about how the unperturbed state transforms under that bare symmetry. Consequently, we show that
the transition between the systems with and without the 1-form symmetry is an \emph{information-theoretic transition} and is analogous to the threshold behavior in quantum error correction (QEC). 
%The error-correction perspective has been exploited to capture emergent properties of the 1-form symmetries in the 3D classical lattice gauge theories~\cite{Self_dual_critically_Ising,Adam_Nahum_2024}. 
The connection between emergent 1-form symmetries and QEC provides a physical foundation for employing QEC-based approaches to detect quantum phases~\cite{QCNN_2019,RG_QEC_exact_2023,LED_2024,Sagar_ViJay_2024, qcnn2d:2024}.

Building on the geometric argument, we propose a precise criterion for the existence of a 1-form symmetry in a quantum state on a lattice system in the thermodynamic limit, even when the symmetry is emergent. The criterion identifies when information about a given bare symmetry remains recoverable and therefore whether a corresponding emergent 1-form symmetry, with a representation independent of the microscopic details, can exist. Using this framework, we derive the transition point for the existence of 1-form symmetries in simple examples of product states (Sec.~\ref{sec:1d_example}). We propose, that the transition from the presence to the absence of a 1-form symmetry is an information-theoretic transition (see e.g. Refs.~\cite{Fan_2024,Jian_2024,AFM_melting_2024,Lee:2025} for examples of such transitions). This information-theoretic transition does not generally lead to singularities in physical observables and therefore does not coincide with a quantum many-body phase transition. However, we demonstrate that this information-theoretic transition manifests as an abrupt change in the long-range entanglement structure of post-measurement states when measurements project the system into specific 1-form symmetry charge configurations. For instance, a short-range entangled state without 1-form symmetry can become long-range entangled after such measurements, connecting measurement-induced long-range entanglement with higher-form symmetries. We formulate a generic theory for emergent 1-form symmetries and discuss its relation to QEC in Sec.~\ref{sec:1form_qec}.

\setcounter{footnote}{0}  
\stepcounter{footnote}

\addtocounter{footnote}{-1} 
\footnotetext{As we work with quantum lattice models with a tensor-product factorized Hilbert space, all bare  1-form symmetries are lattice 1-form symmetries and non-topological~\cite{nat:2020,marvin:2021,oh:2023,choi:2025}, and charged open strings are allowed. For topological 1-form symmetries (say, when restricting to low-energy subspace), charged objects are lines or closed loops without open ends.}

To showcase the applications of our theory, we propose an efficient protocol that (i) detects 1-form symmetries using QEC techniques (Sec.~\ref{sec:detection}) and (ii) probes 2D topological phase transitions characterized by the spontaneous breaking of these 1-form symmetries (Sec.~\ref{sec:detection_topo} and \ref{sec:deformed_states}). The protocol leverages general QEC decoders to detect 1-form symmetries and can be applied to subsystems of a quantum state without requiring specific boundary conditions, which are typically necessary in standard QEC settings.
To probe the topological quantum phase transition, we apply the QEC to restore the bare 1-form symmetry and subsequently perform efficient measurement of disorder parameters that detect the topological quantum phase transitions characterized by the spontaneous breaking of the 1-form symmetry.  
We illustrate the protocol using the Minimum Weight Perfect Matching (MWPM) algorithm~\cite{MWPM_1973,MWPM_1982,MWPM_1986} (albeit any other decoder can be used as well) and examine the emergent 1-form symmetries as well as the topological phase transitions in the perturbed 2D toric code models. We further extend the detection protocol using planar geometries (Sec.~\ref{sec:protocol_for_open_sys}). 
In Sec.~\ref{sec:condensation}, we discuss how our proposed theory offers new insights on the puzzles of particle condensation in the $\mathbb{Z}_2$ lattice gauge theories and show that the Higgs and confinement regime are sharply defined by information theoretic transitions.
We provide an outlook in Sec.~\ref{sec:conclude} and relegate technical details of the calculations to the appendices.

\section{From conventional global symmetries to 1-form symmetries}
A familiar example of symmetry in condensed matter systems is an on-site global symmetry.
Given a group $G$, such a symmetry acts through a unitary representation $u(g)$ that satisfies $u(g)u(h)=u(gh)$ for all $g,h\in G$. An on-site global symmetry is realized as $U(g)=\otimes_i u_i(g)$, where the tensor product runs over unit cells $i$. An operator is symmetric if it commutes with the symmetry operators $U(g)$. Hence, specifying a physical symmetry amounts to choosing a particular representation. 

Fixing (the representation of) symmetries is important and physically meaningful in characterizing many-body systems under symmetries. In a physical phase diagram, the microscopic symmetry action is usually fixed. One then asks, with this symmetry action held fixed, whether the system realizes distinct symmetry-breaking phases, symmetry-protected topological phases~\cite{Pollmann:2010,schuch:2011,chen:2011}, or phase transitions between them (e.g. see concrete examples in~\cite{ruben:2017}). The same fixed symmetry action determines which Hamiltonian perturbations are symmetry-allowed and also guides the definition of the associated order and disorder parameters characterizing the phase transition. 

% Once the symmetry (representation) is fixed, the symmetry constrains the allowed terms in the Hamiltonian as well as guides the definition of the order parameters. Furthermore, the way in which ground states transform under the given symmetry is fundamental for characterizing, for instance, spontaneous symmetry breaking phases or symmetry-protected topological (SPT) phases protected by $G$. 
%For example, a physical system may host several $G$ symmetries, yet a stable non-trivial SPT order is protected only by a subset of them~\cite{verresen:spt,QCNN_Liu_2023}.

% As a simple illustration, consider the ferromagnetic Ising chain $H=\sum_i Z_i Z_{i+1}$, which possesses two $\mathbb{Z}_2$ symmetries generated by $\prod_i X_i$ and $\prod_i Z_i$, where $X$ and $Z$ are Pauli matrices acting on qubit degrees of freedom. The system exhibits $\mathbb{Z}_2$ symmetry breaking only with respect to the first generator, not the second. A small local perturbation that fails to commute with $\prod_i X_i$ but preserves $\prod_i Z_i$ can therefore destabilize the ground-state manifold. Similarly, in SPT phases protected by a symmetry $G$, a physical system may host several $G$ symmetries, yet stable non-trivial SPT order is protected only by a subset of them~\cite{verresen:spt,QCNN_Liu_2023}.

Generalization of this framework to 1-form symmetries immediately introduces new subtleties. For on-site global symmetries, any local operator that fails to commute with $U(g)$ is not symmetric. By contrast, 1-form symmetries can survive weak local perturbations that do not commute with them. The emergent symmetry operators in the  low-energy subspace are therefore dressed by microscopic fluctuations. This distinctive feature raises several questions unique to 1-form symmetries: Under what conditions do 1-form symmetries emerge? When do such symmetries cease to be emergent? And what is the nature of the transition where they disappear?
Addressing these questions will be a crucial first step toward characterizing quantum phases and phase transitions governed by emergent 1-form symmetries.

% Quasi-adiabatic continuation~\cite{Hastings_and_Wen_2005} was used to argue that some 1-form symmetry can still exist within a gapped phase, albeit dressed by the quasi-adiabatic evolution. This framework is particularly useful when considering the existence of anomaly between different symmetries which is preserved under the adiabatic evolution.  
% Nevertheless, we argue that this perspective alone is insufficient for identifying the emergence of a specific symmetry. An adiabatic evolution can, in some cases, act merely as a local basis transformation. As a result, although some 1-form symmetry may continue to exist after such an evolution, its representation can differ drastically from that of the original unperturbed symmetry. Consequently, it becomes unclear whether the information encoded in the unperturbed symmetry representation remains well defined after the evolution.

Over the past decade, higher-form symmetries and their associated anomalies have emerged as a powerful framework for characterizing and understanding topological order. However, it is important to distinguish between topological order itself and the notion of an emergent higher-form symmetry. Topological order is not protected by symmetry and is therefore invariant under arbitrary finite-depth local unitary transformations. Consequently, topologically ordered states related by local basis changes belong to the same phase. Emergent higher-form symmetries, by contrast, concern the existence of specific symmetry structures in the low-energy theory and generally depend on how these structures are represented in terms of microscopic operators. Therefore, while higher-form symmetries provide a useful language for understanding topological order, the two concepts are not equivalent.

In light of this distinction, it is useful to separate our question from the standard notion of dressed symmetries obtained by quasi-adiabatic continuation~\cite{Hastings_and_Wen_2005}, which implies that, along a gapped path starting from a system with microscopic (bare) symmetries, one can always construct dressed symmetry operators acting within the low-energy subspace. These dressed operators generally depend on the microscopic parameters of the path and provide a powerful framework for tracking structures that are invariant under adiabatic evolution, such as symmetry anomalies. Our question, however, is different. We ask whether, despite the explicit breaking of the bare higher-form symmetry at the microscopic level, the low-energy theory nevertheless admits a common emergent higher-form symmetry associated with that same bare symmetry throughout an extended region of parameter space. When such an emergent symmetry exists, it plays a role analogous to a fixed microscopic symmetry: it constrains the low-energy effective Hamiltonian and guides the definition of order and disorder parameters that characterize distinct phases of matter. The key difference is that the symmetry acts only within the low-energy subspace and does not appear as an exact symmetry of the microscopic Hamiltonian.

Indeed, there has been recent indication that the quasi-adiabatic picture alone cannot resolve emergent 1-form symmetries within the topologically trivial phase of the (2+1)D $\mathbb{Z}_2$ theory~\cite{Wen_emergent_high_form_2023}.
Refs.~\cite{Self_dual_critically_Ising,Adam_Nahum_2024} introduced a “patching” algorithm to capture emergent 1-form symmetries in the 3D classical $\mathbb{Z}_2$ lattice gauge theory. Interpreted as a particular error-correction decoder, the algorithm exhibits a transition that numerically matches the expected physical phase transition of the model. However, its applicability beyond classical settings and, in particular, whether the diagnosis depends on the chosen algorithm remains unclear. Therefore, the precise nature of the emergent 1-form symmetries needs to be further investigated. With these considerations in mind, in this work we approach the problem from an information-theoretic vantage point and propose a general criterion of emergent 1-form symmetries in quantum systems. Before we do so in Sec.~\ref{sec:1form_qec}, we illustrate the key challenges for detecting emergent 1-form symmetries with product states in Sec.~\ref{sec:1d_example}.

% summarize our key contributions below:
% \begin{enumerate}
%     \item We propose a general decoder-agnostic definition of emergent 1-form symmetry in quantum systems. We relate the transition of 1-form symmetry to an change of long-range entanglement when measuring the 1-form symmetry operators. 
%     \item The proposed theory reconciles the known properties of emergent 1-form symmetries and provides explanation of the emergent 1-form symmetries complementing the conventional framework of quasiadiabatic continuation. We illustrate this by studying a number of one- and two-dimensional lattice models. 
%     \item As a practical application, we show how the theory inspires efficient protocol for detecting topological quantum phase transition, which is tested against $\mathbb{Z}_2$ topologically ordered systems.
%     \item As a theoretical application, we discuss that the theory provides a general definition for particle condensation, giving rise to a sharply separated Higgs regime and confining regime in $\mathbb{Z}_2$ lattice gauge theories.
% \end{enumerate}

\section{Emergent 1-form symmetries and long-range entanglement: Illustrative example of product states}\label{sec:1d_example}

Before presenting a general characterization for emergent 1-form symmetries, we examine the instructive example of product states. We also use this example to introduce key terminology.
We consider the following Hamiltonian for  $\theta\in[0,\pi/2]$
\begin{equation}\label{eq:h_product}
    H(\theta) = -\sin(\theta)\sum_i X_i - \cos(\theta)\sum_i Z_i,
\end{equation}
with a qubit (i.e. a spin-1/2 degree of freedom) on each site. We use $X,Y,Z$ to denote the Pauli matrices.
The unique ground state of this Hamiltonian is 
\begin{equation}
    \ket{\Phi(\theta)} =\prod_i\ket{\theta}_i \quad \text{with} \quad \ket{\theta}_i=\cos(\theta/2)\ket{0}_i+\sin(\theta/2)\ket{1}_i
    \label{eq:ps_groundstate}
\end{equation}
We will now analyze these states both in one and two spatial dimensions.

\subsection{Product states in one dimension}

Let us consider a one-dimensional (1D) periodic system. A $\mathbb{Z}_2$ 1-form symmetry is generated by taking products of elements from $\{Z_i|\forall i\}$. The resulting set of operators are divided into contractible and non-contractible 1-form symmetry operators. The contractible operators for the symmetry are generated by the set $S=\{Z_iZ_{i+1} |\forall i\}$. A 1-form symmetry operator $Z_iZ_j$ acts on a 0D support and its support can be smoothly deformed by taking a product with elements in $S$. We refer to these 1-form symmetry operators as contractible 1-form symmetry operators as they can be shrunk to identity when $i = j$.  
Furthermore, the operator $Z_i$ for any site $i$ is a distinct 1-form symmetry operator. In contrast to the contractible case, the operators $Z_i$ are non-contractible as they cannot be smoothly deformed to the identity using elements of the set $S$. However, with elements of $S$ the operator $Z_i$ can be deformed from site $i$ to any other site $j$. The 1-form symmetry on a closed manifold is specified by the contractible and non-contractible symmetry operators. A quantum state $\ket{\psi}$ is symmetric under the bare 1-form symmetry if it is an eigenstate of every 1-form symmetry operator.

The charged operators of the 1-form symmetry are line operators which transform non-trivially under the 1-form symmetry operators. In this case, a charged line operator (without open ends) is simply the global Pauli $X$ on all sites that satisfies $(Z_iZ_j)(\prod_k X_k)(Z_iZ_j)^{-1} = \prod_k X_k$ and $Z_i (\prod_k X_k) Z_i^{-1} = -(\prod_k X_k)$. 
This implies that $X$-string operators (segments of line operators with open ends) create 1-form symmetry charges at their endpoints. The contractible symmetry operators of the set $S$ that enclose the endpoints detect the 1-form symmetry charges, as they transform the string operators non-trivially. In the spin language, the 1-form symmetry charges can be interpreted as domain walls generated by the endpoints of $X$-string operators.

When $\theta = 0$, the Hamiltonian and the ground state respect the bare 1-form symmetry generated by $S$. If $\theta\neq 0$, the system no longer has the bare 1-form symmetry but the 1-form symmetry can be emergent, as we investigate now. 

\begin{figure}
    \centering
    \includegraphics{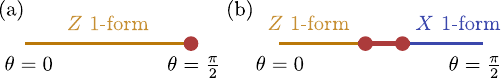}
    \caption{\textbf{1-form symmetries of product states}. (a) In 1D, the  $Z$ 1-form symmetry is generated by $Z_iZ_{i+1}$. Using the criterion we develop, the $Z$ 1-form symmetry is present for $\theta\in [0,\pi/2)$ and disappears at $\theta = \pi/2$. (b) In 2D, the $Z$ 1-form and $X$ 1-form symmetries are generated by $B_p$ and $A_v$, respectively. Via a mapping to the 2D RBIM along the Nishimori line, we find that the $Z$ and $X$ 1-form symmetries exist when $\theta\in[0,\theta_c)$ and $\theta\in(\pi/2-\theta_c,\pi/2]$, respectively, where $\theta_c = 0.22146(2)\pi$. The middle region  $\theta \in [\theta_c, \pi/2-\theta_c]$ shown by the red line has neither the $Z$ 1-form symmetry nor the $X$ 1-form symmetry.}
    \label{fig:productstate}
\end{figure}

To analyze the existence of the emergent 1-form symmetry, we interpret the ground states $\ket{\phi(\theta)}$ of the 1D version of the Hamiltonian in Eq.~\eqref{eq:h_product} as a superposition of different domain-wall configurations, each labeled by its eigenvalues $\pmb{m}$ under the contractible 1-form symmetry operators $Z_iZ_j$. Any given domain-wall configurations can be created in exactly two ways by applying $X$-string operators to $\ket{00\cdots 0}$. For a length-$L$ system, this allows us to write
\begin{align}
    \ket{\phi} &=\sum_{\pmb{m}} \left[c_{\pmb{m},0}\prod_{i\in K}X_i + c_{\pmb{m},1}\prod_{i\notin K}X_i \right]\ket{00\cdots 0},\nonumber\\
    c_{\pmb{m},0} & = \cos^{L-|K|}(\theta/2)\sin^{|K|}(\theta/2),\nonumber \\ c_{\pmb{m},1} &= \sin^{L-|K|}(\theta/2)\cos^{|K|}(\theta/2),
    \label{eq:post_meas}
\end{align}
where $K = K(\pmb{m})$ (with $|K|\leq L/2$) is the set of qubits flipped from $\ket{00\cdots 0}$ to reach the domain-wall configuration $\pmb{m}$. In this picture, we view $\ket{\phi}$ as charged string operators acting on a state symmetric under all the bare 1-form symmetry operators.

Geometrically, the 1-form symmetry emerges from the ability to “deform” the symmetry operators through the charged string operators in Eq.~\eqref{eq:post_meas}, allowing them to commute effectively (see Fig.~\ref{fig:pulling}). For this deformation to be well-defined, each typical charge configuration must be generated by a unique set of charged string operators so that the symmetry can be “pulled through” unambiguously. Using Eq.~\eqref{eq:post_meas}, this condition is
\begin{equation}\label{eq:1d_criteron}
    \lim_{L\to\infty}\sum_{\pmb{m}} \max_{q\in \{0,1\}}\kappa_{\pmb{m},q}
    = 1,\quad \kappa_{\pmb{m},\pmb{q}}  = |c_{\pmb{m},\pmb{q}}|^2.
\end{equation}
If this holds, the 1-form symmetry commutes effectively with the charged string operators and is thus present in the ground state $\ket{\phi}$. Because it is no longer generated by the bare generators in $S$, the symmetry becomes dressed: one specifies how the bare 1-form operators should be deformed through the charged string operators in each charge configuration, and the size of that dressing is the average deformation required.
Conversely, if Eq.~\eqref{eq:1d_criteron} fails, the 1-form symmetry ceases to exist. For $\theta\in[0, \pi/2]$, we have $\kappa_{\pmb{m},0} \geq \kappa_{\pmb{m},1}$ for all $\pmb{m}$, therefore
\begin{equation}
     \lim_{L\to\infty}\sum_{\pmb{m}} \max_{q\in \{0,1\}}\kappa_{\pmb{m},q} = \lim_{L\to\infty} \cos^{2L}(\theta/2)\sum_{i = 0}^{\lfloor L/2\rfloor}\binom{L}{i}\tan^{2i}(\theta/2),
\end{equation}
which evaluates to 1 for $\theta\in [0, \pi/2)$ and evaluates to 1/2 at $\theta = \pi/2$.
Therefore, the 1-form symmetry is lost and has a transition at $\theta = \pi/2$, as illustrated in Fig.~\ref{fig:productstate}a. Physically, below the transition, only a local deformation is required for the bare 1-form symmetry to effectively commute through the charged strings. Under coarse graining, these charged strings become irrelevant at long distances. As a result, the low-energy theory loses sensitivity to the microscopic details of the symmetry-breaking perturbation and admits an emergent 1-form symmetry independent of the microscopic parameters. Indeed, in Appendix~\ref{sec:app:dressZ}, we illustrate an explicit approach to construct the parameter-independent dressed 1-form symmetry operators (for $\theta<\pi/2$) using the idea of renormalization-group circuits.  

A crucial observation is that the Hamiltonian $H(\theta)$ remains gapped for all $\theta$; the loss of the 1-form symmetry is thus not a quantum phase transition. However, the absence of this symmetry coincides with an abrupt change in the entanglement pattern of the post-measurement states once the 1-form symmetry charges are measured and the state is projected onto each charge configuration $\pmb{m}$. When $\theta < \pi/2$, a typical charge configuration satisfies $2|K|/L \approx r <1$ as $L\rightarrow\infty$. Here, $r/2$ is the average number of 1's in a sampled bitstring from the state $\ket{\phi}$. It follows that the projected state converges to a product state $\left(\prod_{i\in K}X_i\right)\ket{00\cdots 0}$. While at $\theta = \pi/2$, the 1-form symmetry is absent and the post-measurement state takes a form similar to the Greenberger–Horne–Zeilinger (GHZ) state with long-range entanglement. Thus, the presence or absence of the 1-form symmetry is reflected by the distinct entanglement structures of the post-measurement states after measuring the contractible generators $Z_iZ_{i+1}$.

Although in this section, the ground state is unique and possesses a product-state structure, the framework we develop applies more broadly to models that do not satisfy these conditions. The proposed criterion Eq.~\eqref{eq:1d_criteron} for the emergent $\mathbb{Z}_2$ 1-form symmetry is applicable to generic 1D systems. For example, it is robust under the action of any local operator; that is, if $\ket{\psi}$ has the emergent 1-form symmetry, then $O\ket{\psi}$ does as well for any local operator $O$. 
%Furthermore, infinitesimal time evolution generated by any local Hamiltonian also preserves the emergent 1-form symmetry. 
As a physical example we consider the 1D quantum Ising model $H = -\sum_i Z_iZ_{i+1} - h_x\sum_i X_i- h_z\sum_i Z_i$. We find that for $h_z=0$ the ground state retains the emergent 1-form symmetry up to the critical point $h_x = 1$; and the emergent 1-form symmetry is not present in the unique ground state for $h_x>1$, consistent with the understanding that the ordered phase results from an anomaly between the 1-form symmetry and the global spin-flip $\mathbb{Z}_2$ symmetry. For finite longitudinal field $h_z>0$, the quantum Ising model has an emergent Z 1-form symmetry for arbitrary $h_x$, consistent with the product-state limit discussed in this section. We present the detailed analysis in Appendix~\ref{sec:Ising_chain}.

\begin{figure}
    \centering
    \includegraphics{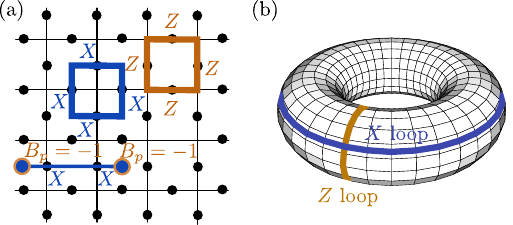}
    \caption{\textbf{Examples of 1-form symmetry operators and charged string operators in 2D.} (a) The $Z$ and the $X$ 1-form symmetries are generated by the plaquette and the vertex operators. The 1-form symmetry charges for the $Z$ 1-form symmetry are created by $X$-string operators. (b) The coexistence of the non-contractible $X$ and $Z$ 1-form symmetry operators implies that a ground-state manifold has topological degeneracy. They cannot coexist at the same time in a trivial product state.}
    \label{fig:noncontractible}
\end{figure}

\subsection{Product states in two dimensions}
\label{sec:2dproduct}
In the following, we extend the reasoning of the previous section to analyze another more non-trivial example of a 1-form symmetry on a 2D square lattice with a qubit at each edge $e$. We consider the system with the periodic boundary condition and the Hamiltonian Eq.~\eqref{eq:h_product} has terms acting on the edges $e$ and has the unique ground state $\ket{\Phi(\theta)}$; see Eq.~\eqref{eq:ps_groundstate}.
At $\theta = 0$, the system has a bare 1-form symmetry whose contractible operators are generated by the set $S = \{B_p = \prod_{e\in p} Z_e|\forall p\}$,
where $p$ denotes the plaquette on the lattice (see Fig.~\ref{fig:noncontractible}a). We refer to the 1-form symmetry as the $Z$ 1-form symmetry. The contractible $Z$ 1-form symmetry operators are obtained by multiplying $B_p$ with different $p$. A non-contractible 1-form symmetry operator is a product of Pauli $Z$ along a non-contractible loop on the torus (see Fig.~\ref{fig:noncontractible}b), which cannot be deformed to identity by taking products with elements in $S$. The charged string operators are Pauli-$X$ strings along the dual lattice.

Following the 1D example in the previous section, a quantum state is said to exhibit a 1-form symmetry if, for a typical 1-form symmetry charge configuration, there exists a unique set of string operators generating those charges such that there is a well-defined way to deform the 1-form symmetry through these charged string operators. In two and higher dimensions, the interior of a charged string operator can be smoothly deformed, unlike the purely endpoint-based deformations in one dimension. From the perspective of a 1-form symmetry, these smoothly deformed operators remain indistinguishable if they yield the same set of eigenvalues when acted upon by all 1-form symmetry operators. For instance, the deformation of the $X$-string operators is achieved by multiplying their interior with the vertex operators $A_v=\prod_{e\in v}X_e$, products of Pauli $X$ around different vertex $v$ (see Fig.~\ref{fig:noncontractible}a), which commute with the 1-form symmetry operators.

This observation naturally extends Eq.~\eqref{eq:1d_criteron} for determining the existence of a 1-form symmetry in the quantum state to the 2D example. For a typical charge configuration labeled by the set of eigenvalues $\pmb{m} = \{m_p\}$ under the contractible 1-form symmetry generators $B_p$ (the set therefore also determines the eigenvalues under all  the contractible 1-form symmetry operators), there should exist a unique set of eigenvalues $\pmb{q}=(q_x=\pm 1,q_y=\pm 1)$ of the non-contractible 1-form symmetry operators $W^{[x]}_Z=\prod_{e\in \mathcal{C}_x} Z_e$ and $W^{[y]}_Z=\prod_{e\in \mathcal{C}_y} Z_e$, where $\mathcal{C}_x$ and $\mathcal{C}_y$ are non-contractible loops on the primal lattice and along the $x$ and $y$ directions. It follows that we can analyze the existence of the 1-form symmetry using:
\begin{equation}  \kappa_{\pmb{m},\pmb{q}}=\bra{\Phi(\theta)}\prod_p\frac{1+m_p B_p}{2}\frac{1+q_x W^{[x]}_Z}{2}\frac{1+q_y W^{[y]}_Z}{2}\ket{\Phi(\theta)},
\end{equation}
which captures the probability of the eigenvalues $\pmb{q}$ for any given 1-form symmetry charge configuration $\pmb{m}$. 
Following this strategy, we find that $\kappa_{\pmb{m},\pmb{q}}$ can be mapped to an instance of the partition function for the 2D random-bond Ising model (RBIM) along the Nishimori line~\cite{Nishimori_1981,Topo_quantum_memory_2002} $\kappa_{\pmb{m},\pmb{q}}\propto\mathcal{Z}_{\text{RBIM}}$, (see Appendix~\ref{app:map_to_RBIM} for the derivation). The mapping reveals a threshold for the existence of the $Z$ 1-form symmetry at a critical angle $\theta_c = 0.2146(2)\pi$, as shown in Fig.~\ref{fig:productstate}b. The regime $0\leq\theta<\theta_c$ corresponds to the ordered phase of the RBIM and different deformation of the charged string operators contributes to the fluctuation of the domain walls in the RBIM, which is suppressed by an energy penalty growing with the size of the domain walls. The macroscopic fluctuation of the charged string operators (for a given $\pmb{m}$) is suppressed and they are equivalent to each other. Therefore, the $Z$ 1-form symmetry is present, i.e. $\lim_{L\to\infty}\sum_{\pmb{m}}\max_{\pmb{q}}\kappa_{\pmb{m},\pmb{q}} = 1$. The regime $\theta_c<\theta\leq \pi/2$ corresponds to the high-temperature disordered phase of the RBIM, the charged string operators are allowed to fluctuate at all scale, $\ket{\Phi(\theta)}$ does not have the $Z$ 1-form symmetry. We reiterate, that the Hamiltonian $H(\theta)$ remains gapped for all $\theta$ and the transition upon losing the $Z$ 1-form symmetry is an information-theoretic transition.

The 1-form symmetry transition is also accompanied by a transition from short-range entangled to long-range entangled post-measurement states upon measuring the 1-form symmetry charges. When $\theta=0$, the projected state is simply the product state $\prod_e\ket{0}_e$ symmetric under the bare $Z$ 1-form symmetry. When $\theta = \pi/2$, the post-measurement states have the form 
 \begin{align}\label{eq:measured_state_plus}
    \ket{\phi_{\pmb{m}}(\pi/2)}\propto \prod_p\frac{1+m_pB_p}{2}\ket{+},
\end{align}
 where $\pmb{m}=\{m_p=\pm 1\}$ is the $Z$ 1-form symmetry charge configuration, and $\ket{+}=(\ket{0}+\ket{1})/\sqrt{2}$. Here, $\ket{\phi_{\pmb{m}}(\pi/2)}$ are nothing but different eigenstates of the toric code model (see Eq.~\eqref{eq:TC_Hamiltonianwithfields} for the definition of the model), each of them is long-range entangled. 
 Slightly away from the limits $\theta = 0$ or $\pi/2$, the measurement projection can be considered as local perturbation to the post-measurement states at those limits, which cannot change the long-range entanglement in the states.
 We expect the same entanglement pattern to persist up to the transition of the 1-form symmetry. We support this with more analysis and numerical simulation in Appendix~\ref{app:Mixed_state_transition}.

We can solve the same problem for another 1-form symmetry generated by $A_v=\prod_{e\in v}X_e$ which we refer to as the $X$ 1-form symmetry. The $X$ 1-form symmetry is, similar to the $Z$ 1-form symmetry, a loop-like symmetry on the dual lattice. The ground state has the bare $X$ 1-form symmetry at $\theta = \pi/2$. The absence of the $X$ 1-form symmetry can be similarly diagnosed by the mapping to the RBIM along the Nishimori line. The ground state of the Hamiltonian $H(\theta)$ therefore has the $Z$ 1-form symmetry for $\theta\in[0,\theta_c)$ and the $X$ 1-form symmetry for $\theta\in (\pi/2-\theta_c, \pi/2]$. In between, the ground state lacks either of the 1-form symmetries, as summarized in Fig.~\ref{fig:productstate}b. If both the $X$ and the $Z$ 1-form symmetries were to coexist in the ground state, then the bare, non-contractible $X$ and $Z$ loops along orthogonal directions, which anticommute, would force the ground-state manifold to transform non-trivially under these loops. This transformation would lead to a degeneracy dependent on the system’s topology, implying the presence of topological order and therefore contradicting the product-state nature of the ground state.

\section{An information-theoretic criterion for the existence of the 1-form symmetry}\label{sec:1form_qec}

The 1D and 2D product-state examples provide an explicit illustration of 1-form symmetries and highlight several key elements underlying their emergence. In this section, we build on the ideas from the previous section to propose a quantitative criterion for the existence of 1-form symmetries in more general settings.
We focus on the 1-form symmetries but a similar picture should hold for $p$-form symmetries with $p>1$. 

For a quantum state $\ket{\psi}$ in $D$ spatial dimensions, a bare 1-form symmetry can be thought of as a set of invertible commuting operators $\{O_i\}$ supported on some closed sub-manifold in $D-1$ dimensions such that $O_i\ket{\psi}=\lambda_i\ket{\psi}$ for all support $i$ and some $\lambda_i\in\mathbb{C}$\footnote{We use the word ``bare'' to refer to the known lattice 1-form symmetry operators and distinguish it from the 1-form symmetry when it is emergent at low energy.}, and smooth deformation of the support of the operators yields another set of operators in $\{O_i\}$. On a closed manifold, the set should include contractible operators which can be shrunk to a point under the smooth local deformation and non-contractible operators which cannot be shrunk to a point. If a quantum state is symmetric under all contractible operators but it is not symmetric under non-contractible operators, the quantum state spontaneously breaks the 1-form symmetry. Then the 1-form symmetry can be recovered by projecting the state into a sector defined by the eigenvalues of the non-contractible operators. The 1-form symmetry charges are created by the endpoints of segments of the line operators charged under the 1-form symmetry. We refer to the charged line operators with open ends as charged string operators.

Following the 1D and 2D examples in Sec.~\ref{sec:1d_example}, given a reference bare 1-form symmetry, a perturbed quantum state \(\ket{\psi}\) is said to exhibit the 1-form symmetry if, for a typical 1-form symmetry charge configuration, there exists a unique set of string operators generating those charges such that there is a well-defined way for the 1-form symmetry to commute through these charged string operators by deformation. In this setting, the action of the 1-form symmetry on the state $\ket{\psi}$ remains well defined, meaning that the way in which the unperturbed quantum state transforms under the bare 1-form symmetry is preserved, even if the bare 1-form symmetry itself is not exactly maintained.
Let $\pmb{m}=\{m_i\}$ denote a charge configuration labeled by the eigenvalues under the bare contractible 1-form symmetry operators $\{O_i\}_{\text{cont}}\subseteq\{O_i\}$. Let $\pmb{q}$ denote the set of eigenvalues under the set of non-contractible bare 1-form symmetry operators $ \{O_i\}\backslash \{O_i\}_{\text{cont}}$. A given quantum state $\ket{\psi}$ can be decomposed into different eigensectors under the bare 1-form symmetry:
\begin{equation}
\label{eq:decompose}
    \ket{\psi} = \sum_{\pmb{m}}\sum_{\pmb{q}} \mathcal{P}_{\pmb{m},\pmb{q}}\ket{\psi},
\end{equation}
where $\mathcal{P}_{\pmb{m},\pmb{q}}$ denotes the projector onto the eigenspace with eigenvalues $\pmb{m}$ and $\pmb{q}$ under the bare 1-form symmetry. For a given charge configuration $\pmb{m}$, the corresponding inequivalent string operators are labeled by different $\pmb{q}$.
For the 1-form symmetry to exist, a given $\pmb{m}$ should have a well-defined $\pmb{q}$ with probability one. 
In the thermodynamic limit, the condition is
\begin{equation}
\label{eq:decodable}\lim_{L\to\infty} \sum_{\pmb{m}}\max_{\pmb{q}} \kappa_{\pmb{m},\pmb{q}}= 1, \quad \kappa_{\pmb{m},\pmb{q}} = \bra{\psi}\mathcal{P}_{\pmb{m},\pmb{q}}\ket
    \psi.
\end{equation}
The loss of the 1-form symmetry corresponds to a transition where Eq.~\eqref{eq:decodable} fails to be satisfied. 
Although we adopt periodic boundary conditions, the presence of a 1-form symmetry is fundamentally a \emph{bulk} property, independent of the boundary conditions. For instance, if the system has open boundaries (where non-contractible operators are not defined), then losing the 1-form symmetry means that the same charge configuration can be generated by macroscopically distinct charged string operators. Henceforth, we say that any state \(\ket{\psi}\) obeying Eq.~\eqref{eq:decodable} is 1-form symmetric, keeping in mind that this symmetry can be emergent.

\subsection{Emergent 1-form symmetries and post-measurement long-range entanglement}
The transition from the presence of a 1-form symmetry to the absence of a 1-form symmetry is generally not a quantum phase transition with singularity in the physical observables measured on the state. Instead, it is an information-theoretic transition
that corresponds to singularities in non-linear functions of the physical states. For instance, consider measuring the 1-form symmetry charges, collapsing the system onto a particular charge configuration $\pmb{m}$. A short-range entangled state lacking a 1-form symmetry can become long-range entangled after such a measurement, meaning it cannot be mapped to a product state via a finite-depth local quantum circuit. This behavior is clearly illustrated by the 1D and 2D examples in Sec.~\ref{sec:1d_example}. A transition in the 1-form symmetry will generally be accompanied by a sudden change of the entanglement in the post-measurement states. The reason is that, in the absence of the emergent 1-form symmetry with non-zero probability, a collapsed charge configuration contains contribution from inequivalent charged string operators. This implies that while the post-measurement state is symmetric under all contractible bare 1-form symmetry operators, 
it is not symmetric under the non-contractible bare 1-form symmetry operators. The post-measurement state spontaneously breaks the 1-form symmetry and is long-range entangled due to the non-local correlation. 

This demonstrates that the transition of a 1-form symmetry can generally be detected by monitoring the change in long-range entanglement of the post-measurement states. However, entropic quantities that capture such entanglement are difficult to access, both in real experiments and in numerical simulations. In the next section, we show that by exploiting the connection to quantum error correction (QEC), one can efficiently detect the presence of 1-form symmetries and approximate the location of the information-theoretic transition.

\subsection{Connection to quantum error correction}
The characterization based on Eqs.~\eqref{eq:decompose}-\eqref{eq:decodable} for the existence of 1-form symmetries is inspired by and intimately related to the decodability problem in QEC~\cite{Topo_quantum_memory_2002}.
The characterization demands that, under perturbation, it is in principle possible to deform the 1-form symmetry operator through the perturbation and thereby allow for a dressed symmetry to emerge. The QEC decoders provide a natural tool to perform such ``deformation.'' Physically, a decoder specifies a  procedure independent of microscopic details for identifying the charged objects associated with the 1-form symmetry operators and removing them. In this sense, the existence of a decoder suggests that the low-energy theory admits an emergent 1-form symmetry independent of microscopic parameters.

When the 1-form symmetry generators of interest are Pauli strings, the decoding technique in the standard QEC directly provides an efficient way to detect the 1-form symmetry. In standard QEC, Pauli strings (i.e. the stabilizers in QEC) play the role of the 1-form symmetry generators. The measurement of the 1-form symmetry charges (the syndrome in QEC) can be efficiently made by measuring the contractible 1-form symmetry generators. 
After collecting the syndrome, a decoder is used to decode the string operators (physical errors in QEC) generating the charge configuration. If for a typical charge configuration $\pmb{m}$, the charged string operators predicted by the decoder has the correct $\pmb{q}$ under the non-contractible 1-form symmetry operators, the QEC is successful and Eq.~\eqref{eq:decodable} is true. We can formulate this more precisely. Consider a decoder that maps a charge configuration $\pmb{m}$ to a particular set of eigenvalues $\pmb{q}$ under the non-contractible 1-form symmetry operators with probability $p(\pmb{q}|\pmb{m})$. We can lower bound the left hand side of Eq.~\eqref{eq:decodable} by

\begin{align}
     \sum_{\pmb{m}}\sum_{\pmb{q}}\kappa_{\pmb{m},\pmb{q}}p(\pmb{q}|\pmb{m})
    \leq \sum_{\pmb{m}}\max_{\pmb{q}} \kappa_{\pmb{m},\pmb{q}}\leq 1. 
\end{align}
If the decoder $p(\pmb{q}|\pmb{m})$ outputs success with probability one, 
\begin{equation}
    \lim_{L\to\infty} \sum_{\pmb{m}} \sum_{\pmb{q}}\kappa_{\pmb{m},\pmb{q}}p(\pmb{q}|\pmb{m}) = 1,
\end{equation}
and the condition Eq.~\eqref{eq:decodable} is fulfilled. The successful QEC is thus an indicator for the existence of the 1-form symmetry. In practice, successful decoding is generally a sufficient but not necessary condition for the existence of the 1-form symmetry. For instance, a decoder could yield a wrong prediction even if $\pmb{q}$ is in principle decodable. An optimal decoder can correctly predict $\pmb{q}$ with probability one as long as the 1-form symmetry exists. In the product state example of Sec.~\ref{sec:1d_example}, we could analytically construct the optimal decoder. For generic cases this is not possible and we will consider approximate decoders, such as Minimum-Weight Perfect Matching (MWPM), instead, which we can efficiently evaluate numerically.

\section{Detection of emergent 1-form symmetries: toric code in a field}\label{sec:detection}

We employ the relation between 1-form symmetries and quantum error correction (QEC) to detect 1-form symmetries for the (2+1)D toric code on the square lattice in a field~\cite{kitaev_2002}. For generic values of the field, one cannot analytically construct the optimal decoder. Instead, we will use MWPM as a concrete decoder.  We argue that our approach extends to a broad class of 2D abelian topological states. Further examples, including deformed (2+0)D toric code states and the double-semion model are presented in Sec.~\ref{sec:deformed_states}. The details for the numerical simulation can be found in Appendix~\ref{app:simulation_method}.

The Hamiltonian of the model reads
\begin{equation}\label{eq:TC_Hamiltonianwithfields}
H(h_x,h_z)=-\sum_{v}A_v-\sum_{p}B_p-h_x\sum_{e}X_e-h_z\sum_{e} Z_e,
\end{equation}
% \begin{equation}\label{eq:TC_Hamiltonian}
% H_{\text{TC}}=-\sum_{v}A_v-\sum_{p}B_p,
% \end{equation}
where $A_v=\prod_{e\in v}X_e$ and $B_p=\prod_{e\in p} Z_e$ are the vertex and plaquette operators, as shown in Fig.~\ref{fig:noncontractible}a. %For zero field, the system has four-fold degenerate $\mathbb{Z}_2$ topologically ordered ground states satisfying $A_v = B_p = +1$, for all $v$ and $p$. For weak finite field, topological order remains robust. A topological phase transition to the paramagnet arises at a critical field. The phase diagram of this model has been extensively studied~\cite{TC_phase_diagram_expansion_2009,Youjin_TC_phase_diagram_2011,TC_XYZ_2011}; see Fig.~\ref{Fig:phase_diagram}(a) for an illustration. 
We focus on the $Z$ 1-form symmetry generated by $B_p$; the protocol targeting the $X$ 1-form symmetry generated by $A_v$ works analogously.\footnote{The toric code model is equivalent to the $\mathbb{Z}_2$ lattice gauge theory. In that context, the $X$ and $Z$ 1-form symmetry operators are Wilson and 't Hooft loop operators.}
The Hamiltonian $H(0,0)$ has both 1-form symmetries. 
The four degenerate ground states of the toric code are labeled by the eigenvalues $(q_x,q_y)$ of the two non-contractible $Z$ loops, $W_{Z}^{[x]}=\prod_{e\in \mathcal{C}_x}Z_e$ and $W_{Z}^{[y]}=\prod_{e\in \mathcal{C}_y}Z_e$, where $\mathcal{C}_y$ and $\mathcal{C}_x$ are two non-contractible loops along the $x$ and $y$ direction of the lattice; see Fig.~\ref{fig:noncontractible}b for an example of $W_{Z}^{[y]}=\prod_{e\in \mathcal{C}_y}Z_e$. 

When $h_x=0$, the ground state $\ket{\psi(h_x=0,h_z)}$ has the bare $Z$ 1-form symmetry, and when $h_z=0$, $\ket{\psi(h_x,h_z=0)}$ has the bare $X$ 1-form symmetry. When $h_x$ and $h_z$ are large enough, a topological phase transition from the topological phase to the trivial phase happens. The phase diagram of this model has been extensively studied~\cite{TC_phase_diagram_expansion_2009,Youjin_TC_phase_diagram_2011,TC_XYZ_2011}. 
When $h_x,h_z\gg 1$, the ground state reduces to the product states considered in Sec.~\ref{sec:2dproduct} with $\theta=\tan^{-1}(h_z/h_x)$.  

\begin{figure}
    \centering
    \includegraphics{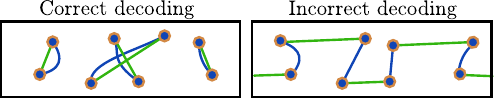}
    \caption{\textbf{Decoding of 1-form symmetry charge configurations.} The Minimum-Weight Perfect Matching (MWPM) decoder takes the 1-form symmetry charge configuration (dots) and outputs the predicted sets of $X$-strings (green lines) required to remove all the $Z$ 1-form symmetry charges created by the actual $X$-strings (blue lines). If the $Z$ 1-form symmetry charges are dense, the algorithm fails by returning a set of recovery $X$-strings (green lines) that is inequivalent to the actual $X$-strings (blue lines).  
    }
    \label{Fig:TC}
\end{figure}

We apply the QEC to the ground states and compare the predicted phase diagram with the one we obtain with our approach.  The protocol is practically implemented as follows: Given a quantum state $\ket{\psi(h_x,h_z)}$, repeat simultaneous measurements of the contractible $Z$ 1-form symmetry generators $B_p$ on all the plaquettes $p$. Feed the sampled  $Z$ 1-form symmetry charge (syndrome) configurations (i.e. $\pmb{m}=\{m_p\}$ where $m_p=\pm 1$ label the eigenvalues of $B_p$) to the MWPM decoder. 
The decoder outputs the sets of shortest Pauli-$X$ strings required to annihilate all the $Z$ 1-form symmetry charges. 
The success of the decoder can be determined by an indicator obtained as follows:  
If the decoder identifies the Pauli-$X$ strings that are equivalent to the string operators creating the $Z$ 1-form symmetry charges, see Fig.~\ref{Fig:TC}, the QEC-recovered states (i.e. the states after the predicted Pauli-$X$ strings are applied to the post-measurement states) are the simultaneous eigenstates of $W_Z^{[x]}$ and $W_Z^{[y]}$ with the eigenvalues $(q_x,q_y)=(1,1)$ in the thermodynamic limit. If this is not the case, then the decoding has failed. An indicator for the presence of the $Z$ 1-form symmetry is thus given by the $\langle W_Z\rangle=\left(\langle W^{[x]}_Z\rangle+\langle W^{[y]}_Z\rangle\right)/2$ from the QEC-recovered states. 
When $\langle W_Z\rangle$ is not $1$ in the thermodynamic limit, then it is unclear whether the 1-form symmetry exists. Only for an optimal decoder we would know that the 1-form symmetry does not exist.

% \begin{figure}
%     \centering
%     \includegraphics[width=1\linewidth]{Fig_main_phase_diagram.pdf}
%     \caption{\textbf{The phase diagram for the toric code model in a magnetic field.} (a) The quantum phase diagram of the system, with the topological and the trivial phase. The dashed lines indicate cuts at which  we have evaluated our protocol, illustrated in Fig.~\ref{Fig:scan_main}. (b) The emergent 1-form symmetries of the system.}
%     \label{Fig:phase_diagram}
% \end{figure}

\begin{figure*}
    \centering
    \includegraphics[scale=0.5]{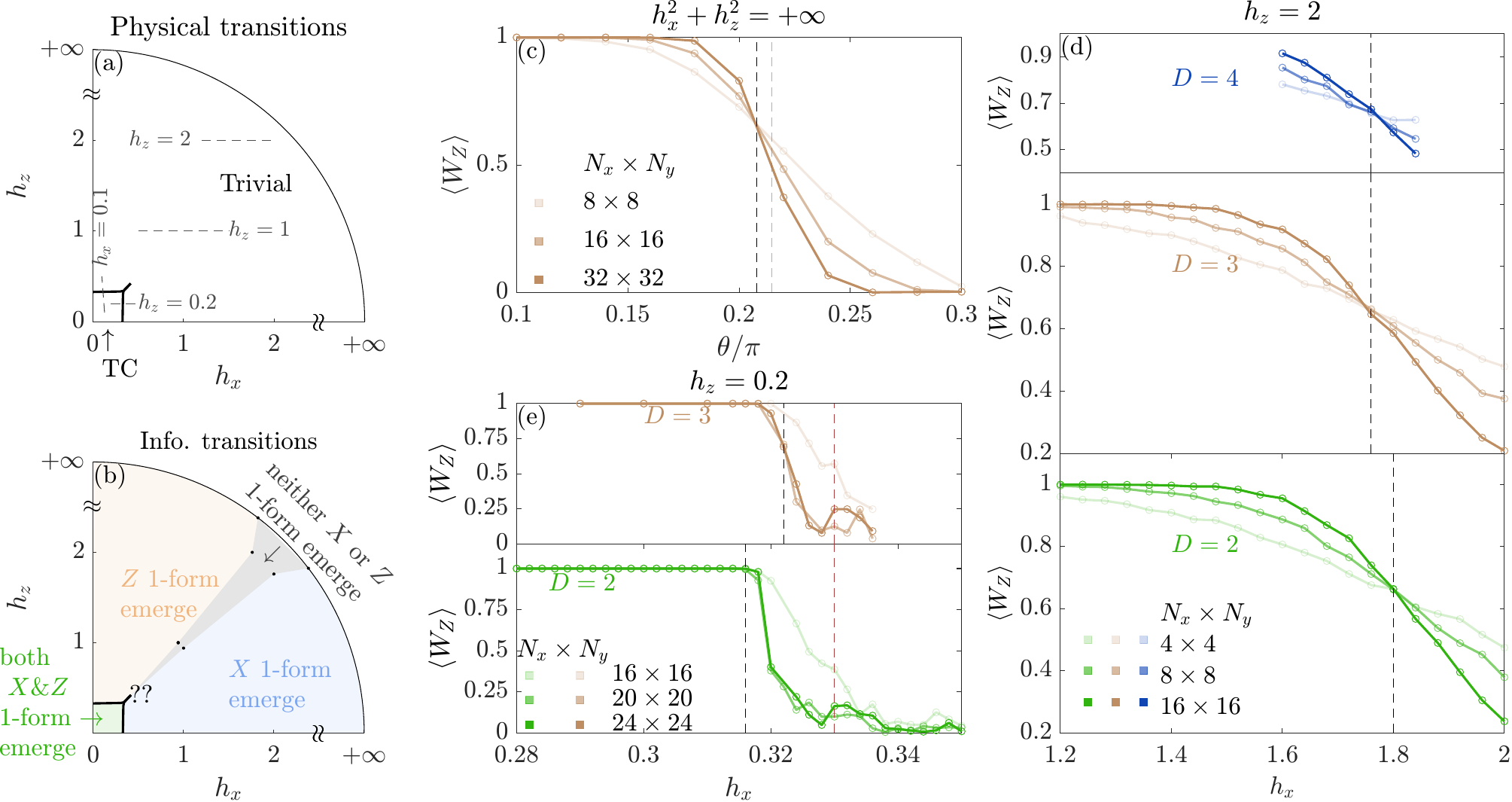}
    \caption{\textbf{Numerical results for detecting $Z$ 1-form symmetry from the ground states of the toric code in a field.} 
   % (a) The theoretical phase diagram of the toric code model in magnetic fields. There are two gapped phases, the TC phase (gree region) and the trivial phase (orange+gray+blue region). The TC phase has both $X$ and $Z$ 1-form symmetries. In the trivial phase, these two 1-form symmetries cannot coexist. Instead, an extended region (gray) lacking either of the 1-form symmetries separates the regions where $X$ or $Z$ 1-form symmetry emerge. 
   We consider the expectation values of the non-contractible $Z$ loop operator $\langle W_Z\rangle$ from the QEC-recovered states as the 1-form symmetry indicator. A value of +1 indicates the existence of the 1-form symmetry. The ground states are approximated by tensor-network states on the torus with bond dimension $D$ and system size $N_x\times N_y$. (a) The quantum phase diagram of the system, with the topological and the trivial phase. The dashed lines indicate cuts at which  we have evaluated our protocol, illustrated in Fig.~\ref{Fig:scan_main}b. (b) The emergent 1-form symmetries of the system. The black dots are the QEC thresholds from the numerical simulation. (c) Loop operator $\langle W_Z\rangle$ in the product-state limit $h_x^2+h_z^2=+\infty$. The black dashed lines indicate theoretical threshold from the MWPM decoder, the green dashed line is the theoretical threshold from the optimal decoder. (d)  Loop operator $\langle W_Z\rangle$ for $h_z=2$. The black dashed lines indicate the crossing points of $\langle W_Z\rangle$ from different system sizes (symbols) and different tensor-network state bond dimensions $D$ (colors).
   (e) Loop operator $\langle W_Z\rangle$ at $h_z=0.2$. Peak of correlation length from the tensor network state at the bond dimension $D$ (dashed black). Location of the topological phase transition (red dashed) obtained from Ref.~\cite{Youjin_TC_phase_diagram_2011}. 
   %The star marks the exact field where the emergent Z 1-form symmetry ceases to exist in the product-state limit. We scan the entire parameter space with a system of $N_x \times N_y = 8\times 8$ plaquettes. Cuts of the phase diagram 
    }
    \label{Fig:scan_main}
\end{figure*}

The quantum phase diagram and the regions with different 1-form symmetries for the ground states are shown in Figs.~\ref{Fig:scan_main}a and b. The information-theoretic transition of the 1-form symmetries is in general independent of the quantum phase transitions. 
In the product-state limit, when $h_x,h_z\gg 1$ (see Sec.~\ref{sec:2dproduct}), there is no quantum phase transition, but there are transitions between the presence and absence of the 1-form symmetries characterized by the RBIM. The gray region in Fig.~\ref{Fig:scan_main}b without the $Z$ or $X$ 1-form symmetry exists away from the product-state limit and terminates at some point along the line $h_x=h_z$ outside the toric code (TC) phase.

% \textcolor{blue}{[Not sure if we have this.]}
% We apply the protocol to numerically scan the information-theoretic transition of the $Z$ 1-form symmetry in the phase diagram of the deformed toric code states. 
% The $Z$ 1-form symmetry indictor $\langle W_Z\rangle$ measured from the QEC recovered states is shown in Fig.~\ref{Fig:scan}b. 
% The information-theoretic transition boundary of the $Z$ 1-form symmetry qualitatively agrees with the theoretical phase diagram Fig.~\ref{Fig:scan}a. The points of the transition in Fig.~\ref{Fig:scan}b are obtained from the MWPM decoder, which is not optimal and underestimates the correct threshold upon which the 1-form symmetry is absent. For this reason, it will not coincide with the phase boundary (blue lines in Fig.~\ref{Fig:scan}a) between the TC phase and the trivial phase.

To benchmark our theoretical prediction of the 1-form symmetries in the product-state limit, we numerically evaluate the Z 1-form symmetry indicator using the MWPM decoder. As discussed in Sec.~\ref{sec:2dproduct}, the $Z$ 1-form symmetry exists when $\theta\in[0,\theta_c)$, where $\theta_c= 0.22146(2)\pi$. 
If an optimal decoder is used, the predicted transition corresponds to the RBIM along the Nishimori line~\cite{Nishimori_1981}. However, the MWPM decoder is not optimal and predicts a lower threshold than the optimal decoder. Interestingly, MWPM decoder corresponds to the RBIM at zero temperature~\cite{Preskill_2003}. It follows that the MWPM decoder correctly identifies the 1-form symmetry up to $\theta_c^{\text{MWPM}}= 0.2081(1)\pi < \theta_c$, consistent with our $Z$ 1-form symmetry indicator $\langle W_Z\rangle$ shown in Fig.~\ref{Fig:scan_main}c. As the system size increases from $8\times 8$ to $32\times 32$ plaquettes, the indicator converges to +1 for $\theta<\theta_c^{\text{MWPM}}$ and to 0 for $\theta > \theta_c^{\text{MWPM}}$. Nonetheless, the MWPM decoder is a reasonably good  decoder in practice for detecting the 1-form symmetries, with a predicted threshold very close to the theoretically optimal threshold.

Away from the product-state limit, the ground states exhibit a finite correlation, which limits the system size of the numerical simulation. To approximate its ground states, we employ variationally optimized 2D tensor-network states (see Appendix.~\ref{app:simulation_method}). We then benchmark MWPM decoder both deep in the trivial phase and near the topological phase transition. Concretely, we first examine the line $h_z = 2$ in the topologically trivial phase. For a given bond dimension, the indicators for different system sizes cross approximately at a single point within the resolution of $h_x$, and we identify this crossing point as the transition point from the MWPM decoder. %We have independently verified that this point does not drift when increasing the system size to $L=32$ and refining the resolution in $h_x$~\cite{In_prapare}. 
The transition point estimated from the MWPM decoder converges as the bond dimension of the tensor network increases (see Fig.~\ref{Fig:scan_main}d), supporting the persistence of the information-theoretic transition beyond the product-state limit.

To probe the 1-form symmetry transition at the vicinity of the topological phase transition, we set $h_z = 0.2$. As the bond dimension increases, the quality of the variational ground states systematically improves, with the peak of the correlation length approaching the correct quantum critical point. Applying the MWPM decoder to these states yields an estimated 1-form symmetry transition that aligns with the correlation-length peak of the variational states (see Fig.~\ref{Fig:scan_main}e). Because of the suboptimality of the decoder, we expect the estimated 1-form transition to lie slightly below the true quantum critical point in the limit of infinitely large bond dimension. While this effect is difficult to resolve in the present case due to numerical accuracy, we demonstrate it explicitly in a simplified 2D model in Sec.~\ref{sec:deformed_states}.

We argue that, in general, the quantum phase transitions induced by condensation of excitations at the endpoints of $X$-strings or the $Z$-strings, including those from the TC phase to the trivial phase, exactly coincide with the loss of the $Z$ or $X$ 1-form symmetry~\cite{Wen_emergent_high_form_2023, Adam_Nahum_2024}, respectively.
%For example, this is the case along the lines $h_z = 0$ or $h_x = 0$, where the critical points between the the TC and the trivial phase are characterized by the condensation of these open-string excitations, which exactly correspond to the 1-form symmetry charges under the two bare 1-form symmetries. 
Along the lines $h_z = 0$ or $h_x = 0$ in the trivial phase where the respective 1-form symmetry charges are condensed, the charged string operators generating them are ``dissolved'' and fluctuate at all scales. In consequence, the corresponding 1-form symmetry is lost. In the TC phase where the 1-form symmetry charges are not condensed and are well-defined, the $X$ and the $Z$ 1-form symmetry coexist. The same argument applies away from the $h_z = 0$ or $h_x = 0$, where now the phase transition is driven by the condensation of the excitations generated by open-ended dressed $Z$ ($X$) 1-form symmetry operators, which create the $X$ ($Z$) 1-form charges.

\section{Detection of 2D topological quantum phase transitions}\label{sec:detection_topo}

A potential application for the detection protocol of emergent 1-form symmetries could be to identify topological quantum phase transitions. However, detecting 1-form symmetries alone may not be the most effective strategy for locating such transitions. First, as we showed above, the loss of a 1-form symmetry is an information-theoretic transition and generally does not coincide with a (topological) quantum phase transition. Second, even when the two transitions do coincide, finding an efficient optimal decoder to pinpoint the transition is in general challenging.
Here, we develop a protocol to accurately determine topological quantum phase transitions, once the 1-form symmetry has been detected.

\subsection{Outline of the protocol}

\begin{figure}
    \centering
    \includegraphics{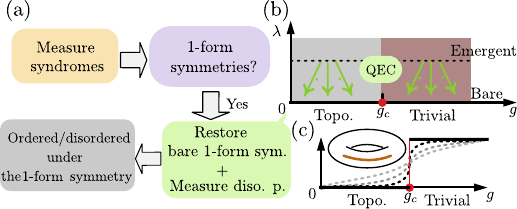}
    \caption{\textbf{Overview of the protocol to detect topological quantum phase transitions}. (a)  The protocol begins by measuring the generators of the 1-form symmetry of the input state, thereby projecting the state into a fixed bare 1-form symmetry charge (syndrome) configuration. These measurements are repeated, and the resulting outcomes are post-processed to compute an indicator for the existence of the 1-form symmetry. If the 1-form symmetry is found to exist, one then corrects for the syndromes by applying the inverse charged string operators [Quantum Error Correction (QEC) step] and measures the disorder parameter of the bare 1-form symmetry QEC-recovered states to distinguish between the ordered and disordered phases of that symmetry. (b) In a gapped system with emergent 1-form symmetry for some $\lambda >0$, the bare 1-form symmetry ($\lambda = 0$) can be restored within the same gapped phase by correcting for the bare 1-form symmetry charges (syndromes) using a standard QEC approach. (c) On the QEC-recovered state with the bare 1-form symmetry, the disorder parameter can efficiently distinguish the topological phase and the trivial phase. In 2D, the disorder parameter is a string operator (inset). 
    }
    \label{Fig:intro}
\end{figure}

A schematic of the protocol is shown in Fig.~\ref{Fig:intro}a. Within our framework, emergent 1-form symmetries are always defined with respect to fixed bare 1-form symmetries. The protocol therefore assumes knowledge of a candidate set of bare 1-form symmetries whose spontaneous breaking gives rise to the topological order of interest.
The protocol first determines whether the emergent 1-form symmetry exists and then probes whether that symmetry is spontaneously broken. More specifically, the full detection protocol consists of two steps:
\begin{enumerate}
    \item \textbf{Detect 1-form symmetries using the protocol of Sec.~\ref{sec:detection}.} 
    \item \textbf{Detect spontaneous breaking of the existing 1-form symmetries.} If the 1-form symmetry is confirmed to exist in Step 1, we recover the bare 1-form symmetry by removing the charges from the post-measurement states based on the decoding in Step 1, as shown in Fig.~\ref{Fig:intro}b. Then 1-form symmetry breaking is detected by, measuring a long bare 1-form symmetry string operator on the QEC-recovered states (Fig.~\ref{Fig:intro}c)~\cite{BRICMONT_1983}. 
\end{enumerate}

\begin{figure}
    \centering
    \includegraphics[scale=0.5]{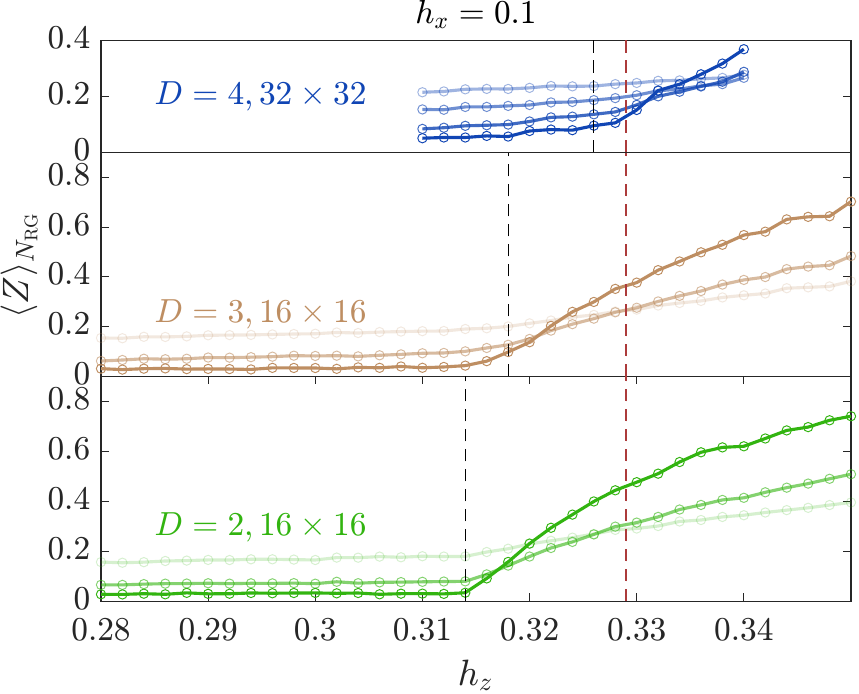}
    \caption{\textbf{Detecting topological phase transition using the disorder parameter from the QEC-recovered ground states of the toric code in a field.} 
    We consider the RG-assisted disorder parameter along the line $h_x=0.1$. The RG applies on the tensor networks states with a bond dimension $D$ put on a torus with the size $N_x\times N_y$. The lines with the same color but from light to dark denotes increasing of the RG steps $N_{\text{RG}}$. The black dashed lines denote the peak of the correlation length of the tensor network state with the bond dimension $D$ and the red dashed line indicates the position of the topological phase transition obtained from Ref.~\cite{Youjin_TC_phase_diagram_2011}. %With increasing iPEPS bond dimension $D$, the topological phase transition is swiftly approached. 
    }
    % (a) The colormap for the expectation values of the RG-assisted disorder parameter $\langle Z\rangle_{{N}_{\text{RG}}}$ with $N_{\text{RG}}=2$ on the QEC-recovered state. The expectation is computed using a single $Z$ operator evaluated after applying 2 steps of the classical RG processing. The detection protocol is applied to the region with the $Z$ 1-form symmetry (see the orange region in Fig.~\ref{Fig:scan}b).
    %  (b) Cut along the line $g_z=0.18$ shown in (a). 
    %  The inset in (b) shows the data collapse using the correlation length critical exponent $\nu=1$ of the 2D classical Ising model, and $g_{z,c}\approx 0.229$. \Wentao{update $D=4$.}}
    \label{Fig:tc_fields_diso}
\end{figure}

In Step 2, the expectation values of the bare 1-form symmetry string operators serve as a disorder parameter for the corresponding bare 1-form symmetry. This disorder parameter can be used to probe topological phase transitions between the 1-form symmetric (disordered or trivial) phase and the phase in which 1-form symmetry is spontaneously broken (topological phase). A nonzero disorder parameter indicates a 1-form symmetric phase, whereas a vanishing disorder parameter signals spontaneous 1-form symmetry breaking.

Unlike the FM string order parameter which is a ratio between exponentially small numbers~\cite{FM_1983,FM_1986}, the disorder parameter on the QEC-recovered states is efficient to measure experimentally, which only requires a sample size independent of the system size in two spatial dimensions. Yet, the disorder parameter measured on the QEC-recovered states is physically equivalent to the FM string order parameter measured on the original quantum state with emergent 1-form symmetry~\cite{xu_2024_FM}, see Appendix~\ref{EC_string_order_and_FM} for detailed analysis.  

Our protocol shows practical advantages compared to previous error-correction inspired techniques to probe topological phase transitions, such as Refs.~\cite{LED_2024,qcnn2d:2024}. These methods adopt a renormalization-group (RG) mindset and therefore restrict attention to \emph{local} decoders implemented through successive coarse-graining steps.  Within our framework, the decoders merely approximate the regime in which an emergent 1-form symmetry exists.  Our protocol advances beyond these techniques in two key respects. (a) We show that the presence of an emergent 1-form symmetry is set by a \emph{global}, information-theoretic criterion; locality of the decoder is unnecessary.  This insight permits the use of global QEC decoders, which possess significantly higher thresholds than local RG-type decoders (see Appendix~\ref{app:MWPM_vs_local_decorder} for an explicit numerical comparison) and thus determine the emergent symmetric regime with much greater accuracy. (b) We demonstrate that detecting the emergent 1-form symmetry alone, which implicitly is the strategy adopted in prior work, often fails to identify the true transition point because of decoder sub-optimality.  Our protocol circumvents this limitation by employing a global decoder to construct an effective FM order parameter that precisely captures the topological phase transition and its universality.  Detection of the emergent symmetry then serves as an internal consistency check validating the applicability of the method. 

It is also worth noting that although the protocol assumes knowledge of the bare symmetry operators, these operators need not be known exactly. One may equally well define the disorder parameter with respect to a set of symmetry operators obtained from the bare ones by a finite-depth local unitary circuit $U$. We expect the resulting detection of the physical quantum phase transition to remain robust, provided each local gate in $U$ is sufficiently close to the identity. The reason is that evaluating the rotated disorder parameter for a Hamiltonian, say $H(h_x,h_z)$ in Eq.~\eqref{eq:TC_Hamiltonianwithfields}, is equivalent to evaluating the original disorder parameter for the rotated Hamiltonian $U^\dagger H(h_x,h_z)U$. Consequently, the robustness of the rotated disorder parameter reduces to the robustness of the original disorder parameter under a small local rotation of the Hamiltonian. For sufficiently weak local rotations, the ground state is only locally modified, producing a dilute density of symmetry charges or errors relative to the original bare-symmetry representation. As long as the system remains below the decoding threshold, these errors can be removed by the QEC decoder. Equivalently, the rotated Hamiltonian remains within the regime exhibiting the same emergent 1-form symmetry as the original system, where the disorder parameter is expected to be effective. Therefore, the rotated disorder parameter likewise remains robust and continues to accurately identify the phase transition.

%In practice, one usually has some basic knowledge about the system, which is typically sufficient to make a reasonable initial guess for the relevant bare 1-form symmetries, even away from a fixed-point limit. Should such an educated guess not be available,a variational algorithm, see for example Ref.~\cite{cian2022extracting} could be used to find a good approximation of the reference bare 1-form symmetry to implement the protocol.  %This variational search is in principle costly, but it needs to be performed only for a handful of representative models chosen strategically, after which our protocol can diagnose emergent symmetries and corresponding topological phase transitions across an entire family of states under study.

\subsection{Results for the toric code model in a field}
\label{sec:deformtc_topodetect}

We now apply our protocol to the ground states in Sec.~\ref{sec:detection}. Concretely, the steps are as follows: if the $Z$ 1-form symmetry is detected, we measure its generators, then restore the bare $Z$ 1-form symmetry of the post-measurement state by removing the 1-form symmetry charges via the decoded charged string operators (Pauli-$X$ strings). Next, we projectively measure the recovered state in the computational (Pauli-$Z$) basis. The resulting measurement outcomes are then processed and averaged to obtain a disorder parameter (a Pauli-$Z$ string much longer than the correlation length of the system) that distinguishes between the phase in which 1-form symmetry is spontaneously broken and the 1-form symmetric phase.

To accurately locate the phase transition in a finite system, one typically performs a careful finite-size scaling analysis (e.g., by computing a Binder cumulant). To streamline this process, we introduce a classical renormalization-group (RG) algorithm for coarse-graining the measured outcomes. The transition point can be identified as the crossing of the RG-assisted disorder parameters at different RG steps. Technical details on implementing the RG-assisted $Z$-string operator can be found in Appendix~\ref{app:QCNN}.

We evaluate the RG-assisted disorder parameter $\langle Z\rangle_{{N}_{\text{RG}}} $ from the QEC-recovered states to identify the topological phase transitions, where $N_{\text{RG}}$ is the number of applied RG transformations. In the region with the $Z$ 1-form symmetry (see Fig.~\ref{Fig:scan_main}b), the phase boundary between the TC phase and the trivial phase is accurately determined by the change from $\lim_{N_{\text{RG}}\rightarrow\infty}\langle Z\rangle_{{N}_{\text{RG}}} =0$ in the TC phase to $\lim_{N_{\text{RG}}\rightarrow\infty}\langle Z\rangle_{{N}_{\text{RG}}} =1$ in the trivial phase. 
We now analyze the topological phase transition for $h_x = 0.1$. In this case, the $Z$ 1-form symmetry is always present, but a topological quantum phase transition from the TC phase to the trivial phase occurs as $h_z$ increases. As shown in Fig.~\ref{Fig:tc_fields_diso}, clear signatures of the transition are visible at each finite bond dimension $D = 2, 3$ and $4$. The predicted critical points (the onset of a non-zero disorder parameter) for increasing $D$ align well with the peaks of the correlation length, which themselves converge toward the true quantum critical point of the system. 

Within our framework, the RG-assisted disorder parameter further captures the universality of the transition. Although the resolution is limited in this example, we demonstrate consistent evidence of universality extraction in a simplified 2D model in Sec.~\ref{sec:deformed_states}.
We emphasis that the quality of the results is not limited by our protocol, but it is limited by the accuracy of the tensor network states approximating the ground states, which is a complicated task as we discuss in Appendix.~\ref{app:simulation_method}.

\begin{figure}
    \centering
    \includegraphics[scale=0.5]{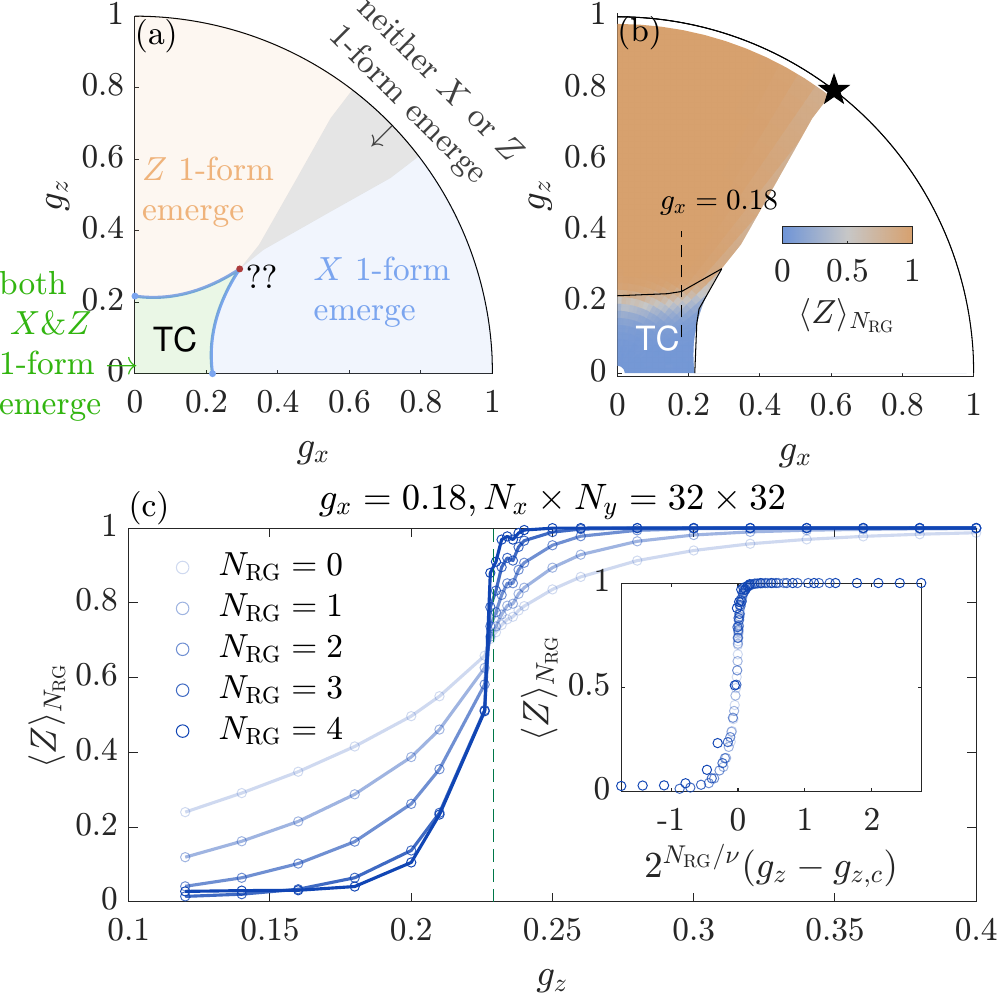}
    \caption{\textbf{Numerical results for the deformed toric code states.} (a) The quantum phases and 1-form symmetries of the deformed toric code states. The green region is the topological TC phase, while the gray, orange and blue regions are the non-topological phase. (b) The detection protocol is applied to the region with the $Z$ 1-form symmetry [green and orange regimes in (a)]. The colormap for the expectation values of the RG-assisted disorder parameter $\langle Z\rangle_{{N}_{\text{RG}}}$ with $N_{\text{RG}}=2$ on the QEC-recovered state. The expectation value is computed using a single $Z$ operator evaluated after applying 2 steps of the classical RG processing. The results are analogous for the regime with $X$ 1-form symmetry [green and blue regimes in (a)]. (c) A zoom-in view of the RG-assisted disorder parameter along the cut $g_x = 0.18$. The black and green dashed lines (which overlap well in the plot) label the predicted and the theoretical transition point, respectively. The inset shows the data collapse using the correlation length critical exponent $\nu=1$ of the 2D classical Ising model, and $g_{z,c}\approx 0.229$.}
    \label{fig:deformTC}
\end{figure}

\section{Benchmarking the accuracy of the protocol with deformed toric code states in (2+0)D}\label{sec:deformed_states}

So far, we have applied the protocol to the ground states of the (2+1)D toric code in magnetic fields and qualitatively consistent results have been obtained. Nevertheless, the numerical accuracy is limited by the finite bond dimensions of the tensor-network states in the simulation. To benchmark the protocol with higher accuracy, we illustrate our protocol on a family of deformed tensor-network ground states that share a similar phase diagram with the ground states in Sec.~\ref{sec:detection} (see Fig.~\ref{fig:deformTC}a). Yet they are exactly described by a 2D tensor-network state with a finite bond dimension, which allows for an analytical treatment in certain regimes and high-accuracy numerical simulation. 

We consider the deformed toric code states~\cite{Gauging_quantum_state_2015,haegeman2015shadows,Zhu_2019}:
\begin{equation}\label{eq:deformed_wavefunction}
    \ket{\psi(g_x,g_z)}=\prod_e(1+g_x X_e+g_z Z_e)\ket{\tTC},
\end{equation}
where $\ket{\tTC}$ is a ground state of $H(0,0)$ in Eq.~\eqref{eq:TC_Hamiltonianwithfields} labeled by the eigenvalues $(q_x,q_y)=(1,1)$ of $W_{Z}^{[x]}$ and $W_{Z}^{[y]}$, and $g_x$ and $g_z$ are tuning parameters satisfying $g_x>0,g_z>0$ and $g_x^2+g_z^2\leq1$. Similar to the toric code in the fields, when $g_x$ and $g_z$ are large enough, a topological phase transition from the topological phase to the trivial phase happens. 
When $g_x^2+g_z^2=1$, Eq.~\eqref{eq:deformed_wavefunction} reduces to the product states considered in Sec.~\ref{sec:2dproduct} with $\theta=\tan^{-1}(g_z/g_x)$. 

In Sec.~\ref{sec:detection}, we argued that the information-theoretic 1-form symmetry transition out of the topologically ordered phase coincides with the condensation transition of the particles generated by the open 1-form symmetry operators. Indeed, by mapping the wavefunction amplitudes to the weights in partition functions of 2D classical Ising models with different boundary conditions~\cite{Sagar_ViJay_2024} and applying the criterion Eq.~\eqref{eq:decodable}, we show explicitly that along $g_x= 0$ or $g_z = 0$, the $Z$ and the $X$ 1-form symmetry coexist in the TC phase up to the critical point, and one of the 1-form symmetries is lost beyond the critical point (see Appendix~\ref{app:mapping_g_z=0}). 
In analytically intractable regime, we apply the protocol to numerically scan the information-theoretic transition of the $Z$ 1-form symmetry in the phase diagram of the deformed toric code states. The detected emergent 1-form symmetries are summarized in Fig.~\ref{fig:deformTC}a (see Appendix~\ref{app:sec:deformTC} for more details of the numerics).
We scan around the physical phase transition using the phase transition detection protocol in Sec.~\ref{sec:detection_topo},  the RG-assisted disorder parameter is shown in Fig.~\ref{fig:deformTC}b. For higher resolution, we zoom in along the line $g_x = 0.18$. The location of the transition detected by the RG-assisted disorder parameter agrees well with the theoretical transition point, as shown in Fig.~\ref{fig:deformTC}c. 
We find that a data collapse of the order parameter is consistent, up to the system sizes we reach, with the 2D Ising university class that describes the topological phase transition of the deformed toric code wave function, see the inset of Fig.~\ref{fig:deformTC}c.

Ground states with distinct topological order can be characterized by different 1-form symmetries. In such cases the detection of an emergent 1-form symmetry can be used to get useful information about the topological phases as well. As an example, we consider the double-semion model~\cite{Freedman:2004,Levin_Wen_2005} which has $\mathbb{Z}_2$ topological order twisted by non-trivial 3-cocycles~\cite{Dijkgraaf_Witten_1990,Levin_Gu_2012}. The TC phase and the double-semion phase have the same topological entanglement entropy~\cite{TEE_levin_Wen_2006,TEE_Kitaev_Preskill_2006}, making them hard to distinguish from this perspective. Nevertheless, we can use the symmetry detection protocol to distinguish the TC phase from the double-semion phase using the fact that only the TC phase can have the $X$ 1-form symmetry coexisting with the $Z$ 1-form symmetry but the double-semion phase does not. We illustrate that the symmetry detection can be used to distinguish the two phases by implementing the detection protocol to a family of deformed quantum states realizing a topological phase transition between the TC phase and the double-semion phase~\cite{Xu_2018}. The protocol indeed correctly predicts extended region in the phase diagram with TC and double-semion topological order (see Appendix~\ref{sec:tctods} for an illustration of our results).

\section{Modified detection protocol for emergent 1-form symmetries in planar geometries}\label{sec:protocol_for_open_sys}

\begin{figure}
    \centering
    \includegraphics{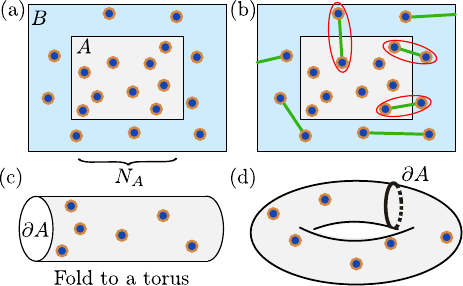}
    \caption{\textbf{Detecting 1-form symmetries from a subsystem $A$.} 
    The illustration uses a cylinder but it works analogously for a disk-like subsystem. (a) A decoder is run on the entire system $A+B$. (b) Remove the charges connected by strings across the boundary between $A$ and $B$. (c) Fold and connect the top and the bottom boundaries of the subsystem $A$ to a cylinder. The folded system has boundary $\partial A$. (d) Fold to a torus by connecting $\partial A$. Run the decoder on the folded torus and compare the new recovery strings in $A$ and the original recovery strings in $A$ across the connected boundary $\partial A$.
    }
    \label{Fig:protocol_for_subsystem}
\end{figure}

In Sec.~\ref{sec:detection} and Sec.~\ref{sec:detection_topo}, the protocol for detecting the 1-form symmetry is implemented on 2D quantum states with periodic boundary conditions. Furthermore, the quantum (ground) states need to be adiabatically connected to a state with the prescribed eigenvalues under the non-contractile 1-form loop operators. 
However, the existence of 1-form symmetries in a quantum state is a bulk property which should be detectable without assuming the boundary conditions. In this section, we propose a modified detection protocol which can be applied to arbitrary disk-like subsystems in an input quantum state. The modified protocol is therefore amenable to various experimental setups.

Consider a system with arbitrary boundary conditions divided into subregions $A$ and $B$ (see Fig.~\ref{Fig:protocol_for_subsystem}a). We show how to modify the protocol in Sec.~\ref{sec:detection} to the subsystem $A$ of a quantum state. 
\begin{enumerate}
    \item Measure the syndrome (1-form symmetry charges) of the entire system, and solve the recovery string (charged string operators) configuration $\mathcal{S}_{\pmb{m}}$ using the a decoder from a given syndrome configuration $\pmb{m}$. This can be done for a system with any boundary condition. For instance, in the toric code model with an open boundary condition, the decoder is allowed to connect a syndrome to the boundary.
    
    \item The recovery strings given by the decoder occur in three types:  $\mathcal{S}_{\pmb{m}}=\mathcal{S}_{\pmb{m},A}\cup \mathcal{S}_{\pmb{m},A\cap B}\cup \mathcal{S}_{\pmb{m},B}$. Each string in $\mathcal{S}_{\pmb{m},A}$ ($\mathcal{S}_{\pmb{m},B}$) only connects a pair of charges that are both in $A$ ($B$), and each string in $\mathcal{S}_{\pmb{m},A\cap B}$ connects one charge in $A$ and the other charge in $B$. Suppose the charge configuration in $A$ is $\pmb{m}_A$, we remove charges in $A$ which are the endpoints of strings in $\mathcal{S}_{\pmb{m},A\cap B}$, as shown in Fig.~\ref{Fig:protocol_for_subsystem}b. 
    
    \item Perform the following classical processing:
    Place the classical syndrome configuration $\pmb{m}'_A$ on a torus formed by folding the boundary of the subsystem $A$, see Figs.~\ref{Fig:protocol_for_subsystem}c and d. Run the decoder on this artificial torus geometry. Measure and average the parity of the number of times that $\mathcal{S}'_{\pmb{m}_A',A}$ crosses $\partial A$, see Fig.~\ref{Fig:protocol_for_subsystem}d. 
\end{enumerate}

The last step above provides a modified indicator for the existence of 1-form symmetries.
We test the modified indicator using the ground states of the Hamiltonian Eq.~\eqref{eq:TC_Hamiltonianwithfields} as well as the deformed toric code states where the average parity of the number of times that $\mathcal{S}'_{\pmb{m}_A',A}$ crosses $\partial A$ can be expressed as the expectation value $\langle \tilde{W}_Z\rangle$ evaluated on the states $\prod_{e\in \mathcal{S}'_{\pmb{m}_A',A}} X_e\ket{00\cdots0}$ on the artificial torus. In Appendix~\ref{sm:sec:modifiedind}, we numerically demonstrate the effectiveness of the protocol with several models, including the 2D perturbed toric code models in the previous section.   
 
Once we confirm the existence of the 1-form symmetry using the modified indicator, we can proceed to detect 2D topological quantum phase transitions either on the subsystem $A$ or on the entire system, as done in Sec.~\ref{sec:detection_topo}. 

% Previous QEC-based approaches for detecting quantum phases either rely on RG-type algorithms which are QEC decoders with a local structure~\cite{QCNN_2019,RG_QEC_exact_2023} or require careful control of the boundary condition~\cite{Sagar_ViJay_2024}. 
The modified indicator described in this section shows how a general global QEC decoder can be applied to subsystems of quantum states. 
This approach provides a general recipe to improve other QEC-based approaches by using a better performing global decoder. 
For example, in Appendix~\ref{sm:sec:modifiedind}, we show that the modified indicator based on a optimal global majority-vote decoder can indeed accurately determine the phase transition between 1D $\mathbb{Z}_2\times \mathbb{Z}_2$ SPT phases whose ground states have a structure similar to that of a 1-form symmetry.  

\section{Emergent 1-form symmetries and particle condensation}\label{sec:condensation}

Our result is not only useful in practical scenarios, it also serves as a new tool for understanding physical phenomena. 
As an example, we show how our framework offers a new perspective on a long-standing theoretical problem of sharply defining the Higgs regime in the $\mathbb{Z}_2$ lattice gauge theory.
In most gauge theories, the Higgs (where the gauge bosons are massive and the charges are condensed) and confined (where the static charges cost extensive energy to separate) phases are expected to be distinct, corresponding to different physical mechanisms of mass generation and confinement. However, in $\mathbb{Z}_2$ lattice gauge theories, Fradkin and Shenker showed that these two regimes are smoothly connected without a phase transition~\cite{Fradkin:1979}, as shown by the phase diagram in Fig.~\ref{Fig:scan_main}a. This is counterintuitive because the two regimes are described by very different microscopic physical pictures. 

In the context of condensed matter systems, the model of the toric code in magnetic fields describes a $\mathbb{Z}_2$ lattice gauge theory with dynamical matter fields~\cite{TMC_Kitaev_2010,xu_2024_entanglement,Xu_Roughen_2025}, where the electric charges or (matter degrees of freedom) are the vertex excitations ($e$-anyon) and the magnetic charges are the plaquette excitations ($m$-anyon). In the limit $h_z = 0$ and $h_x$ is varied, the electric charges undergo a deconfinement-confinement transition across the topological phase transition, i.e., the energy between two separated vertex excitations is constant in one phase and grows linearly with the separation in the other phase~\cite{Wilson_confinement_1974}. Confinement arises from the condensation of the magnetic charges. In the TC phase, the ground states are vacua without any magnetic charges created by $X$-strings. In the trivial phase, the vacuum becomes a condensate of magnetic charges characterized by a non-zero expectation value of the $X$-string operators. A non-trivial mutual statistics between the magnetic and the electric charges then causes the confinement of electric charges in this condensate. The opposite happens when $h_x = 0$ and $h_z$ is varied, the condensation of the electric charges causes the confinement of the magnetic charges, which we refer to as the Higgs regime in connection to lattice gauge theories. 

This analysis of particle condensation relies on a crucial assumption that at least one 1-form symmetry is exactly preserved, i.e., $h_x=0$ or $h_z=0$. When $h_z$ and $h_x$ are both non-zero, the conventional criterion used for diagnosing confinement (the perimeter/area law of the Wilson loops~\cite{Wegner_duality_1971,Kogut_1979}) or condensation (the disorder parameter of the 1-form symmetry~\cite{BRICMONT_1983}) fails. It therefore remains an open question how the Higgs and confining regimes are generally distinct from each other.

Our proposed theory offers a resolution from an information-theoretic perspective by providing a precise definition of the regimes where the physical picture of the particle condensation remains valid, even away from the fine-tuned limits. Specifically, by leveraging our 1-form symmetry criterion, the confining regime can be rigorously characterized as the portion of the trivial phase in which the $X$ 1-form symmetry is present and the corresponding disorder parameter remains nonzero. Here, the disorder parameter is defined by a long $X$-string whose endpoints support magnetic charges; its nonvanishing nature reflects the condensation of magnetic charges. The requirement of the $X$ 1-form symmetry ensures that the (dressed) long $X$-strings are creation operators for the magnetic charges at the endpoints. A similar approach precisely defines the Higgs regime, where electric charges condense, based on the $Z$ 1-form symmetry. Consequently, the confining and Higgs regimes, each featuring condensation of magnetic and electric charges, respectively, are separated by an intermediate regime lacking both $X$ and $Z$ 1-form symmetries which is neither the Higgs nor confining regime (see Fig.~\ref{Fig:scan_main}b). As we show in this work, the boundaries between these three regimes are sharply determined by the underlying information-theoretic transition, but do not exhibit signatures in the physical phase diagram (see Fig.~\ref{Fig:scan_main}a). %It would be interesting to compare these viewpoints with earlier related proposals~\cite{Ruben_Higgs_SPT,xu_2024_entanglement} 
Although the MWPM decoder is suboptimal and the intermediate regime is expected to shrink under an optimal decoder, we expect the intermediate regime to persist when perturbing away from the $h_x^2 + h_z^2 =+\infty$ limit.~\footnote{We offer two supporting arguments: First, MWPM turns out to be close to optimal also at weak fields. The evidence is that at the topological phase transition, the MWPM threshold is very close to the exact critical point, as shown in Fig.~\ref{Fig:scan_main}e and Fig.~\ref{Fig:scan}c. suggesting that results from an optimal decoder would not differ substantially from MWPM in this regime. Second, when perturbing away from the product-state limit, the perturbed state remains adiabatically connected to the product state. So the syndrome distribution $P_{\pmb{m}}$ deforms smoothly. Consequently, away from the optimal threshold, feeding this smoothly varying distribution into an optimal decoder is not expected to abruptly change the decoding behavior.} 

Additionally, since the 1D Ising chain subject to both transverse and longitudinal fields can be viewed as a $(1+1)$D lattice gauge theory~\cite{Gauging_Kitaev_2021}, our theory can be applied to determine the Higgs and confining regimes in the 1D case as well. In the absence of a longitudinal field, the ferromagnetic phase may be understood as arising from a mixed anomaly between an emergent $Z$ 1-form symmetry (whose bare generators are local $Z$ operators) and the 0-form spin-flip symmetry generated by $\prod_i X_i$. This is directly analogous to the mixed anomaly between the $Z$ and $X$ 1-form symmetries of the 2D toric code. As the transverse field is increased through the Ising critical point, the emergent $Z$ 1-form symmetry disappears, and the system enters the paramagnetic phase, where the anomaly is absent, see Appendix~\ref{sec:Ising_chain} for details. %The condensations of the charges generated by Pauli $Z$ (associated with the bare $Z$ 1-form symmetry in Sec.~\ref{sec:1d_example}) causes the otherwise deconfined domain walls (generated by products of Pauli $X$) to be confined, which emulates the $\mathbb{Z}_2$ lattice gauge theory in 1D (see Appendix~\ref{sec:Ising_chain}). 

\section{Conclusion and outlook}\label{sec:conclude}

In this work, we provide an information-theoretic perspective and a precise criterion for the emergence of 1-form symmetries, emphasizing its close connection to the decoding problem in quantum error correction (QEC). We posit that the emergence of a 1-form symmetry is fundamentally tied to recoverability: perturbations may dress the bare 1-form symmetry operators by creating charged excitations, but these excitations remain decodable, allowing the original symmetry action to be recovered. From this perspective, QEC decoders provide a natural tool for characterizing emergent 1-form symmetries. Physically, the existence of a decoder throughout an extended region of the phase diagram indicates that the low-energy theory admits a realization of a common emergent 1-form symmetry within this entire region, independent of microscopic details. This emergent symmetry then plays a role analogous to that of a microscopic symmetry, providing the appropriate order and disorder parameters for characterizing distinct phases of matter, even when the microscopic symmetry are explicitly broken.

%reflecting the preservation of information about the bare 1-form symmetry. More precisely, the existence of a QEC decoder

Motivated by this connection, we propose a protocol for detecting the existence of 1-form symmetries from the bulk of a quantum ground state, which in turn enables efficient probing of 2D topological quantum phase transitions in experiments. While our primary focus is on how 1-form symmetries manifest in ground states, the same information-theoretic criterion could be used to probe 1-form symmetries in the eigenstates of an Hamiltonian at various energies. %i.e., some high energy eigenstates with confined excitations~\cite{Xu_Roughen_2025}. 
It would be especially intriguing to pinpoint a critical energy scale below which the 1-form symmetry persists and above which it is lost.  
Moreover, we stress that the characterization for emergent 1-form symmetry in Eq.~\eqref{eq:decodable} depends only on the spectral data of the symmetry operators. Hence, our framework applies beyond bare 1-form symmetries generated by Pauli strings. For example, one can consider the 1-form symmetries whose operators are commuting matrix-product operators (MPOs) with finite bond dimension~\cite{sahinoglu_characterizing2021}. A prominent family of such cases arises in Abelian string-net models (such as the double-semion model)~\cite{Levin_Wen_2005}: each bare 1-form symmetry corresponds to a closed string operator associated with an elementary anyonic excitation. A systematic study of these generalizations will be crucial for a complete understanding of generic 1-form symmetries, and we leave this exciting direction for future work. 

Although our focus is on 1-form symmetries, the condition Eq.~\eqref{eq:decodable} naturally extends to general higher-form symmetries with $p>1$. For example, while 1-form symmetries are generically unstable at finite temperature, 2-form symmetries in systems with $D>1$ spatial dimensions can be thermally stable when their charges, which are now created by sheet-like operators, cost finite energy that is at least proportional to the perimeter of the sheet-like operators. By the standard Peierls argument, energetics dominate over the entropic contribution at sufficiently low temperature, the fluctuation of the sheet-like operators will be suppressed and the state satisfies a condition similar to Eq.~\eqref{eq:decodable}, Consequently, low-temperature thermal fluctuations preserve 2-form symmetry~\cite{Wen_emergent_high_form_2023}. Furthermore, the proposed general connection between the transition in emergent higher-form symmetries and long-range entanglement in the post-measurement states might provide a new possible framework for characterizing measurement-induced long-range entanglement and criticality~\cite{lee:2022,nat:2023,Eckstein:2024} or mixed-state phases. 

A particularly interesting directions enabled by our framework is the potential of a new paradigm for emergent symmetry-protected topological (SPT) %and symmetry-breaking 
phases. These are distinct quantum phases of matter that cannot be adiabatically connected when the symmetry is preserved. The conventional concept of ``symmetry'' is only well defined when the symmetries are exactly preserved. This begs the question of whether there exist, e.g. distinct SPTs protected by emergent higher-form symmetries~\cite{Ruben_Higgs_SPT,xu_2024_entanglement}. This has remained largely unexplored due to the lack of a suitable theoretical framework for emergent 1-form symmetries. Extending the SPTs to emergent symmetries removes the fine-tuned character of these quantum phases; the question is therefore both theoretically and practically relevant.

Furthermore, equipped with a concrete framework, it can be explored under which conditions the information-theoretic transition of 1-form symmetries aligns with a quantum many-body phase transition; an alignment that would be particularly valuable for efficiently detecting and characterizing various quantum phases. For instance, we anticipate that quantum phase transitions characterized by the condensation of the 1-form symmetry charges generally coincide with the 1-form symmetry transition, as argued and illustrated in Sec.~\ref{sec:detection}.

Another open question is how to generalize our criterion and detection protocol to non-abelian topological states with emergent \emph{non-invertible} higher-form symmetries~\cite{Xu_2020,Xu_2022}, which would require more complex, potentially “fusion-aware” quantum error correction schemes~\cite{QEC_Fib_string_net_2022}. Recent work demonstrates that topological quantum phase transitions between symmetry-enriched topological phases can be efficiently realized on a quantum computer~\cite{Lukas_2023,liu_2024_isoTNS, Boesl2025}, suggesting that it would be worthwhile to extend QEC-based protocols systematically to more general symmetry-protected phases by leveraging connections to higher-form symmetries.

\section{Acknowledgment}
%\Wentao{Add deformed TC and modified indicator to appendix after we converge.}
We thank Sagar Vijay, \"Omer Aksoy and Xiao-Gang Wen for the helpful discussions. We are especially grateful to Salvatore Pace for providing valuable feedback on the manuscript. W.-T. Xu would like to thank H.-K. Jin for providing the code for compressing matrix product states. We acknowledge support from the Deutsche Forschungsgemeinschaft (DFG, German Research Foundation) under Germany’s Excellence Strategy--EXC--2111--390814868, TRR 360 – 492547816 and DFG grants No. KN1254/1-2, KN1254/2-1, the European Research Council (ERC) under the European Union’s Horizon 2020 research and innovation programme (grant agreement No. 851161 and No. 771537), the European Union (grant agreement No 101169765), as well as the Munich Quantum Valley, which is supported by the Bavarian state government with funds from the Hightech Agenda Bayern Plus. YJL also acknowledges support from MIT Center for Theoretical Physics (NSF Challenge Institute for Quantum Computation Award No. 10434).

% \noindent\textbf{Competing interests.---}
% The authors are inventors on a patent (EP 24 204 212.5) dated 2 October 2024 that covers the algorithms for detecting topological order. The authors declare no other competing interests.

\noindent\textbf{Data availability.---}
Data and codes are available upon reasonable request on Zenodo~\cite{zenodo}.

\appendix
\addtocontents{toc}{\string\tocdepth@munge}

\section{Constructing the dressed 1-form symmetry operators}\label{sec:app:dressZ}

As mentioned in the main text, a 1-form symmetry emerges when it effectively commutes with the charged string operators. Physically, the emergent 1-form symmetry operators are “dressed,” with a size determined by the average deformation needed to move the bare 1-form symmetry operators through the charged string operators. In principle, one can construct these dressed 1-form symmetry operators by specifying how the bare 1-form symmetry operators act on each possible configuration of 1-form symmetry charges. In this section, we illustrate another way to understand the dressed 1-form symmetry operators by using renormalization-group (RG) circuits, also known as quantum convolutional neural networks (QCNN)~\cite{QCNN_2019,RG_QEC_exact_2023}.

\begin{figure}
    \centering
    \includegraphics[width=1\linewidth]{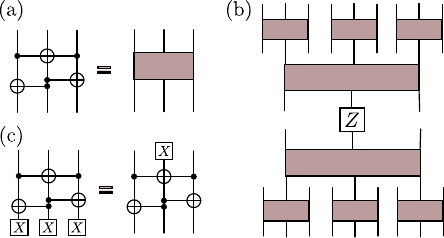}
    \caption{\textbf{Constructing dressed 1-form symmetry operators for the 1D product state example.} The circuit takes input at the bottom and ouput the sate at the top. (a) The scale-invariant component of the RG circuit performs a majority vote on three qubits, setting the middle qubit to 1 if the input bitstring has more 1s than 0s, and to 0 otherwise. The two-qubit gate is CNOT, which flips the target qubit if the control qubit (black dot) is $\ket{1}$. The three-qubit gate is CCNOT, which flips the target qubit if both control qubits (black dots) are $\ket{1}$.  (b) The dressed 1-form symmetry operators are constructed by conjugating the bare 1-form symmetry operators with the RG circuit. (c) The relation implies that the dressed 1-form symmetry has the same charged line operators as the bare 1-form symmetry operators.}
    \label{fig:app:dressZ}
\end{figure}

Consider the 1D product-state example in Sec.~\ref{sec:1d_example}. A useful strategy for constructing the dressed 1-form symmetry is to perform a real-space coarse-graining on the lattice. In doing so, the length of the charged string operators—here, $X$-string operators—can be reduced until they become short enough that the bare 1-form symmetry requires only minimal deformation to pass through them on the coarse-grained lattice. From the analysis in Sec.~\ref{sec:1d_example}, we know that $X$-strings can be optimally decoded from the 1-form charge (domain-wall) configuration via a majority-vote procedure. We can then build an RG circuit that repeatedly applies a local majority vote in a scale-invariant manner. 

Fig.~\ref{fig:app:dressZ}a shows the scale-invariant component of this RG circuit. It acts on three input qubits, with the middle output qubit encoding the majority-vote result. This middle qubit thus represents the coarse-grained version of the three input qubits. Repeating this process $l$ times in a scale-invariant way reduces a system of size $3^M$ to one of size $3^{M-l}$ encoded in the middle-qubit subsystem. After each RG step, the $X$-string operators that create domain walls on the coarse-grained subsystem become shorter by a factor of 1/3. For a system with finite-length $X$-string operators, the final coarse-grained subsystem obtained after these steps will be nearly domain-wall free, meaning it is almost invariant under the bare 1-form symmetry operator $Z_i Z_j$. The dressed 1-form symmetry operators can then be found by conjugating the bare 1-form symmetry operators with this RG circuit, as illustrated in Fig.~\ref{fig:app:dressZ}b. The size of the dressing on these new 1-form symmetry operators is therefore fixed by the average length of the $X$-string operators in the original state. Approaching $\theta = \pi/2$, the average length of the $X$-string operators, which determines the size of the dressing, increases and eventually diverges at $\theta = \pi/2$. From Fig.~\ref{fig:app:dressZ}c, one can also see that these dressed 1-form symmetry operators are independent of the microscopic parameters and retain the same charged line operators as the bare 1-form symmetry, and thus share the same macroscopic properties.

A similar approach can be used to construct the dressed 1-form symmetry operators for the 2D example in Sec.~\ref{sec:1d_example}, for instance by employing the circuit-based RG decoders of Refs.~\cite{LED_2024,Timothy_2024}. A caveat is that these RG decoders in 2D are no longer strictly optimal. However, we expect they can be made arbitrarily close to optimal by incorporating increasingly sophisticated RG steps near the threshold.

\section{The $Z$ 1-form symmetry of 2D product states}\label{app:map_to_RBIM}

In this section, we show that for 2D product states, the distribution of the charged string operators characterized by the eigenvalues $(\pmb{m}, \pmb{q})$ under the $Z$ 1-form symmetry is described by the RBIM. The ground state of the Hamiltonian is given in Eq.~\eqref{eq:ps_groundstate}. The product state is also equivalent to the deformed toric code state in Eq.~\eqref{eq:deformed_wavefunction} at $g_x^2+g_z^2=1$ with $\theta=\tan^{-1}(g_x/g_z)$.

To find the decomposition Eq.~\eqref{eq:decompose}, we use that the eigen-subspace projectors for the $Z$ 1-form symmetry are generated by a product of the commuting local projectors of $B_p$. In other words, we measure the state $\ket{\Phi(\theta)}$ using $B_p, \forall p$. Suppose the measurement outcome, which specifies a charge configuration, is $\pmb{m}=\{m_p\}$, where $m_p=\pm 1$ is the eigenvalue of $B_p$. The corresponding post-measurement state (without normalization) can be written as:
\begin{align}\label{eq:measured_state}
    &\prod_p\frac{1+m_pB_p}{2}\ket{\Phi(\theta)}\notag\\
    &=\sum_{\mathcal{S}_{\pmb{m}}}\left(\sin\frac{\theta}{2}\right)^{|\mathcal{S}_{\pmb{m}}|}\left(\cos\frac{\theta}{2}\right)^{N_e-|\mathcal{S}_{\pmb{m}}|}\prod_{e\in \mathcal{S}_{\pmb{m}}}X_e\ket{00\cdots0},
\end{align}
where $\mathcal{S}_{\pmb{m}}$ is the $X$-string configuration and the endpoints of $X$ strings are at the plaquettes $\{p|m_p=-1\}$, $|\mathcal{S}_{\pmb{m}}|$ is the total length of the $X$ strings, $N_e$ is the total number of edges of the lattice. The $X$ strings are the charged string operators for the $Z$ 1-form symmetry and live on the dual lattice.
The probability of the measurement outcome (charge configuration) can be expressed as:
\begin{align}\label{eq:syndrome_prob}
    &P_{\pmb{m}}=\bra{\Phi(\theta)}\prod_p\frac{1+m_p B_p}{2}\ket{\Phi(\theta)}\notag\\
    &= \sum_{\mathcal{S}_{\pmb{m}}}\left(\sin\frac{\theta}{2}\right)^{2|\mathcal{S}_{\pmb{m}}|}\left(\cos\frac{\theta}{2}\right)^{2N_e-2|\mathcal{S}_{\pmb{m}}|}=\sum_{\mathcal{S}_{\pmb{m}}}p^{|\mathcal{S}_{\pmb{m}}|}(1-p)^{N_e-|\mathcal{S}_{\pmb{m}}|},
\end{align}
where $p = \sin^2\left(\theta/2\right)$. We want to evaluate the charge resolved probability
\begin{equation}
    \kappa_{\pmb{m},\pmb{q}}=\bra{\Phi(\theta)}\prod_p\frac{1+m_p B_p}{2}\frac{1+q_x W^{[x]}_Z}{2}\frac{1+q_y W^{[y]}_Z}{2}\ket{\Phi(\theta)},
\end{equation}
where $\pmb{q}=(q_x=\pm 1,q_y=\pm 1)$ and $W^{[x]}_Z$ and $W^{[y]}_Z$ are two non-contractible $Z$ loop operators on the primal lattice. Notice that $P_{\pmb{m}}=\sum_{\pmb{q}}\kappa_{\pmb{m},\pmb{q}}$. 
To evaluate $\kappa_{\pmb{m},\pmb{q}}$, we essentially need to evaluate Eq.~\eqref{eq:syndrome_prob} restricted to particular $\pmb{q}$. This can be achieved by first picking a particular $X$-string configuration $\mathcal{S}_{\pmb{m},\pmb{q}}$ that is compatible with $(\pmb{m},\pmb{q})$ (More precisely, define an operator $S=\prod_{e\in \mathcal{S}_{\pmb{m},\pmb{q}}}X_e$, it should satisfy $B_pSB^{\dagger}_p=m_pS$, $W^{[x]}_ZSW^{{[x]}\dagger}_Z=q_xS$ and $W^{[y]}_ZSW^{{[y]}\dagger}_Z=q_yS$). Any other $X$-string configuration that is equivalent to the chosen $X$-string configuration has the same $(\pmb{m}, \pmb{q})$ and can be obtained by a smooth deformation of the strings. This implies that the summation over all the equivalent $X$ string configurations in Eq.~\eqref{eq:syndrome_prob} can be performed by introducing virtual Ising spins ($\sigma_v = \pm 1$) at each vertex of the lattice and interpret the other equivalent $X$-string configurations as a result of the domain-wall fluctuation of the virtual Ising spins. 
For instance, to account for configurations with no 1-form symmetry charges and $q_x = q_y = 1$, we can choose the reference $|\mathcal{S}_{\pmb{m}, \pmb{q}}| = 0$. Any $X$-string deformable to the empty configuration have weights 
determined by the Boltzmann weights of the domain-wall configurations formed by the $X$-strings, e.g.,
\begin{equation}\label{app:eq:domain}
    \includegraphics[scale = 1]{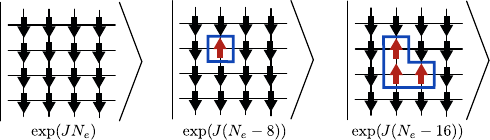},
\end{equation}
where $N_e$ is total number of lattice edges, the blue lines are $X$-loops and the arrows denote the virtual spins at the vertices. Summing over all virtual spin configurations is equivalent to summing over all $X$-strings deformable into one another. In the end, we need to divide this sum by 2 because each domain-wall configuration arises from two opposite virtual spin configurations. For the particular $X$-string configuration $\mathcal{S}_{\pmb{m},\pmb{q}}$ that is compatible with the measurement $(\pmb{m},\pmb{q})$, the Ising couplings are inverted on the bonds of the reference $X$-string $\mathcal{S}_{\pmb{m},\pmb{q}} = \{s_e\}$ configurations. The Ising coupling thus becomes random, and can be written as $J(1-2s_e)$, with $s_e= 1$ if $e\in\mathcal{S}_{\pmb{m},\pmb{q}}$ and $s_e = 0$ otherwise.
For instance, a more typical $X$-string configuration contains end points, the resulting different domain-wall configurations look like
\begin{equation}
    \includegraphics[scale = 1]{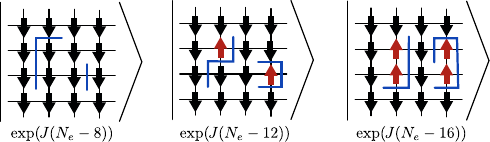}.
\end{equation}

For sectors with non-contractable loops or open $X$-strings, the $X$-strings cannot be deformed to an empty $X$-string configuration. However, the mapping can be carried out analogously  when flipping the signs of the magnetic coupling $J \to -J$ on bonds that overlap with the reference $X$-strings encoded as
$\mathcal{S}_{\pmb{m},\pmb{q}} = \{s_e\}$. Subsequently, we obtain the Boltzmann weights from the virtual spin configurations by deforming the domain walls
\begin{equation}
    \includegraphics[scale = 1]{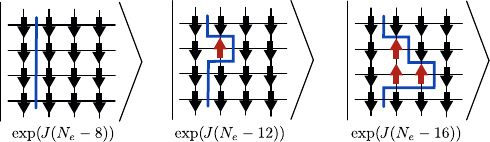}.
\end{equation}

Using this mapping, $\kappa_{\pmb{m},\pmb{q}}$ can be rewritten as the partition function $\mathcal{Z}_{\text{RBIM}}(J,\mathcal{S}_{\pmb{m},\pmb{q}})$ of the RBIM~\cite{Topo_quantum_memory_2002,Sagar_ViJay_2024}:
\begin{align}\label{eq:RBIM}
   \kappa_{\pmb{m},\pmb{q}}&=\frac{[p(1-p)]^{N_e/2}}{2}\sum_{\{\sigma_v\}}\exp\left[ J\sum_{\langle vev'\rangle}(1-2s_e)\sigma_v\sigma_{v'}\right]\notag\\
   &=\frac{[p(1-p)]^{N_e/2}}{2}\mathcal{Z}_{\text{RBIM}}(J,\mathcal{S}_{\pmb{m},\pmb{q}}),
\end{align}
where the syndrome-independent prefactor $[p(1-p)]^{N_e/2}$ has been introduced such that $\kappa_{\pmb{m},\pmb{q}}$ is a probability. In this expression, the Ising coupling is related to the probability as $\exp(-2J)=p/(1-p)$, which is the Nishimori line in the RBIM~\cite{Nishimori_1981}. We emphasize that for a given $(\pmb{m},\pmb{q})$, there are many choices for the reference configuration $\tilde {\mathcal{S}}_{\pmb{m},\pmb{q}}$, but all choices compatible with $(\pmb{m},\pmb{q})$ give rise to the same $\mathcal{Z}_{\text{RBIM}}(J,\tilde{\mathcal{S}}_{\pmb{m},\pmb{q}})$. The fluctuations of the equivalent $X$-strings correspond to the fluctuations of the effective domain walls satisfying $(1-2\tilde s_e)\sigma_v\sigma_{v'} = -1$. 

The next step is to evaluate Eq.~\eqref{eq:decodable} using the partition function of RBIM. We replace the sum $\sum_{\pmb{m}}$ with the sum  $\sum_{\mathcal{S}}$ where $\mathcal{S}=\{s_e\}$ enumerates all the possible $X$-string configurations; this yields:  
\begin{align}\label{eq_app:cal_eq_2}
    \lim_{N_x,N_y \rightarrow \infty} &\sum_{\pmb{m}} \max_{\pmb{q}} \kappa_{\pmb{m}, \pmb{q}} \notag =\lim_{N_x,N_y \rightarrow \infty} \sum_{\pmb{m}} P_{\pmb{m}} \frac{\max_{\pmb{q}} \kappa_{\pmb{m}, \pmb{q}}}{\sum_{\pmb{q}} \kappa_{\pmb{m}, \pmb{q}}} \notag \\
    &= \sum_{\mathcal{S}} \Big[\prod_e p^{s_e} (1-p)^{1-s_e}\Big] \frac{\text{max}_{\hat{\mathcal{L}}}\mathcal{Z}(J,\mathcal{S}+\hat{\mathcal{L}})}{\sum_{\hat{\mathcal{L}}}\mathcal{Z}(J,\mathcal{S}+\hat{\mathcal{L}})},
\end{align}
where $\hat{\mathcal{L}}=\{\emptyset,\hat{\mathcal{C}}_x,\hat{\mathcal{C}}_y,\hat{\mathcal{C}}_x\cup \hat{\mathcal{C}}_y,
\}$ with $\hat{\mathcal{C}}_x$ ($ \hat{\mathcal{C}}_y$) being a non-contractible loop along the $x$ ($y$) direction of the dual lattice.
With this mapping, we can use the same argument in Ref.~\cite{Topo_quantum_memory_2002} to deduce the transition threshold, which we reproduce here for completeness.
We expect that, the free energy difference 
$-\log[\mathcal{Z}(J,\mathcal{S}+\hat{\mathcal{L}})/\mathcal{Z}(J,\mathcal{S})]$ diverges with the linear system size in the ferromagnetic phase of RBIM and it remains a constant in the paramagnetic phase of RBIM, when $\hat{\mathcal{L}}\neq\emptyset$. Eq.~\eqref{eq_app:cal_eq_2} is therefore 1 (smaller than 1) in the ferromagnetic (paramagnetic) phase of the RBIM. Since the parameters of the RBIM are fixed to the Nisimori line $\exp(-2J)=p/(1-p)$ we find that the loss of the (emergent) $Z$ 1-form symmetry in the product states are exactly described by the phase transition from the FM to the paramagnetic phase of the RBIM along the Nishimori line~\cite{Nishimori_1981}.

For the MWPM decoder we use in the numerical part of this work, the detected threshold of the product states corresponds to the RBIM at zero temperature~\cite{Preskill_2003}, i.e., $J=+\infty$. Since the phase transitions of the RBIM along the Nishimori line and at the zero temperature happen at $p_{c}=0.1094(2)$~\cite{Position_of_Nishimori_2001} and $p_{c}^{\text{MWPM}}= 0.1031(1)$~\cite{Preskill_2003}, we can derive the corresponding critical angles $\theta_{c}=0.2146(2)\pi$ and $\theta_c^{\text{MWPM}}=0.2081(1)\pi$. So the optimal threshold and the threshold measured using MWPM are distinct but close to each other.

\section{Post-measurement states of the 2D product states}\label{app:Mixed_state_transition}

We argue that in the 2D product-state limit, a typical $B_p$ post-measurement state is topologically ordered when there is no $Z$ 1-form symmetry. When $\theta=0$, measuring $B_p$ does not change the product state, according to Eqs.~\eqref{eq:measured_state} and ~\eqref{eq:syndrome_prob}. When $\theta=\pi/2$, the post-measurement states are equal weight superpositions of eigenstates of the fixed point toric code model, which can be transformed to the ground states of the fixed point toric code model by applying Pauli strings, so all post-measurement states at $\theta=\pi/2$ have the same long-range entanglement (or topological order) as the ground states of the toric code model.

Then we want to show that a typical post-measurement state remains topologically ordered even slightly away from the limit $\theta=\pi/2$. 
Let us consider an instance of the post-measurement states in Eq.~\eqref{eq:measured_state},
it can be further expressed as:
\begin{align}
    &\ket{\Phi_{\pmb{m}}(\theta)}=\prod_p\frac{1+m_pB_p}{2}\prod_e\ket{\theta}_e\notag\\
    &\propto\prod_{e\in\mathcal{\mathcal{S}_{\pmb{m}}}}X_e\prod_p\frac{1+B_p}{2}\prod_{e\in\mathcal{\mathcal{S}_{\pmb{m}}}}X_e\prod_e\left[1+\tan(\frac{\pi}{4}-\frac{\theta}{2})Z_e\right]\ket{+}_e\notag\\
    &=\prod_{e\in\mathcal{\mathcal{S}_{\pmb{m}}}}X_e\prod_p\frac{1+B_p}{2}\prod_{e}\left[1+(-1)^{s_e}\tan(\frac{\pi}{4}-\frac{\theta}{2})Z_e\right]\ket{+}_e\notag\\
   &=\prod_{e\in\mathcal{\mathcal{S}_{\pmb{m}}}}X_e\prod_{e}\left[1+(-1)^{s_e}\tan(\frac{\pi}{4}-\frac{\theta}{2})Z_e\right]\ket{\tTC},
\end{align}
where $\mathcal{S}_{\pmb{m}}$ is a string configuration which is consistent with a given syndrome configuration $\pmb{m}$. In the last step we used that $B_p$ commutes with $Z_e$.
Since the topological order in $\ket{\tTC}$ is robust against the perturbations on physical degrees of freedom, all $\ket{\Phi_{\pmb{m}}(\theta)}$ have topological order provided $\theta$ is sufficiently close to $\pi/2$ (i.e., the operator $\prod_{e\in\mathcal{\mathcal{S}_{\pmb{m}}}}X_e$ is a constant-depth quantum circuit which does not change the topological order). Thus, we expect that almost all post-measurement states are topologically ordered when $\theta\in(\pi/2-\theta_c,\pi/2)$ in the thermodynamic limit. 

\begin{figure}
    \centering
    \includegraphics[scale=0.5]{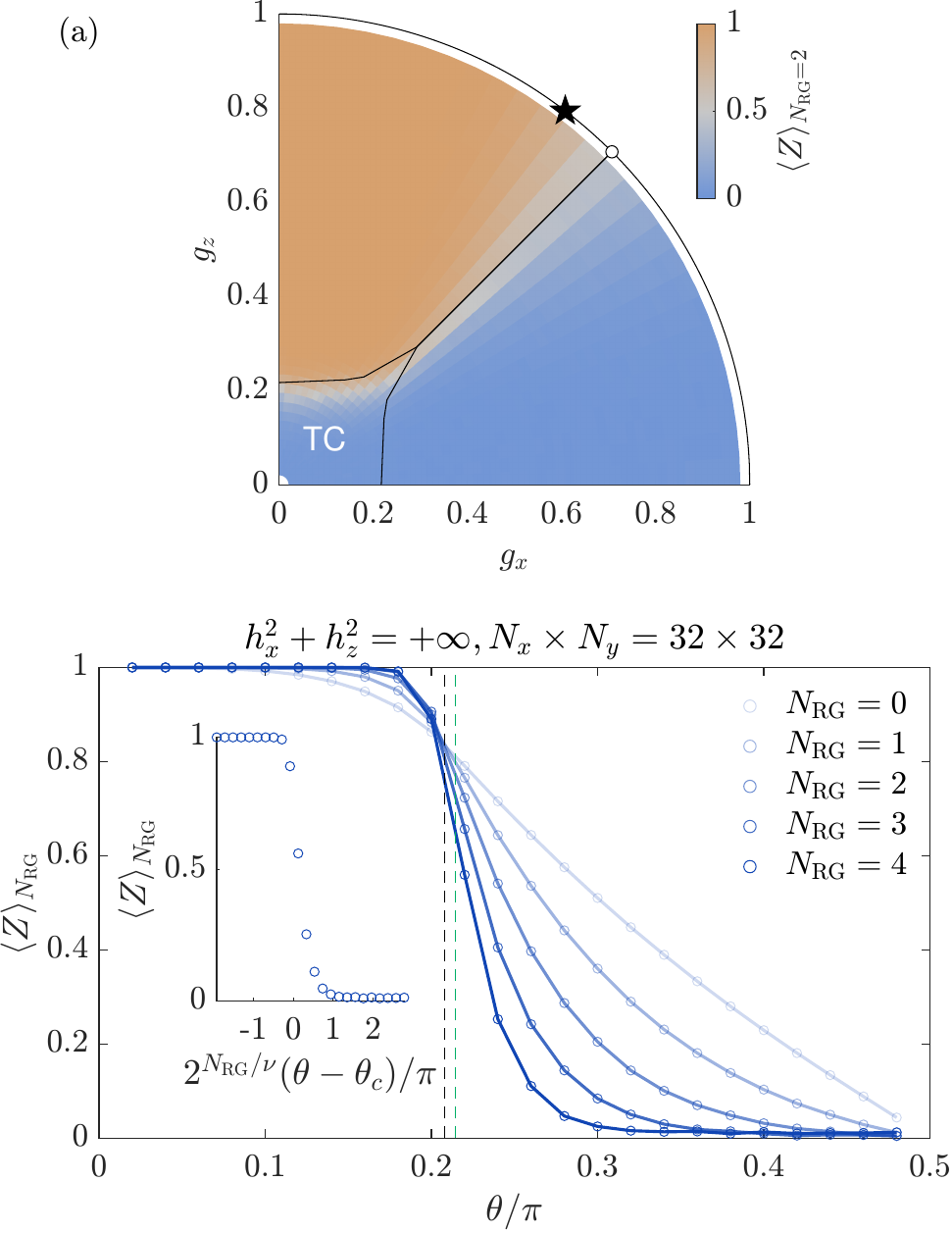}
    \caption{\textbf{Detecting the transition in the post-measurement states using the RG-assisted disorder parameter along the product-state limit $h_z^2+h_x^2=+\infty$. %in Sec.~\ref{sec:detection_topo}.
    } 
    %(a) The colormap for the expectation values of the RG-assisted disorder parameter $\langle Z\rangle_{{N}_{\text{RG}}=2}$ on the QEC-recovered state. The expectation is computed using a single $Z$ operator evaluated after applying 2 steps of the classical RG processing. 
     %(b) 
      The black and green dashed lines mark the numerical and the theoretical transition points, respectively. The inset  shows a data collapse using the correlation length critical exponent $\nu\approx 1.49$ of the 2D RBIM at zero temperature, and $\theta_c=0.208$. %\Wentao{I suggest here we only keep (b) and we do not need (a).}
     }
    \label{Fig:app:product}
\end{figure}

We provide numerical evidence of the transition by detecting a topological phase transition with the method in Sec.~\ref{sec:detection_topo}. 
%We compute the RG-assisted disorder parameter across the entire phase diagram of the deformed toric code states (i.e. an extended version of Fig.~\ref{Fig:inf_field}a), as depicted in Fig.~\ref{Fig:app:product}a. 
 %We can get a better resolution by zooming into the product-state limit. 
 The results in the product-state limit are shown in Fig.~\ref{Fig:app:product}. A transition from +1 to 0 is observed at $\theta_c^{\text{MWPM}}$. %which transition is observed even away from the physical topological quantum phase transition. 
 This transition signifies the transition from short-ranged entangled post-measurement states to long-ranged entangled post-measurement states. From the analysis in Sec.~\ref{sec:2dproduct}, when $\theta\in( \theta_c^{\text{MWPM}}, \theta_c) $ we expect the restored state to remain disordered under the 1-form symmetry and therefore $\langle Z\rangle_{N_{\text{RG}}}$ tends to one.  However, the simulation suggests $\langle Z\rangle_{N_{\text{RG}}}<1$ in the thermodynamic limit.  The behavior can be attributed to the incorrect decoding of the recovery Pauli $X$ strings by the MWPM decoder, the expectation values of the RG-assisted disorder parameters are non-zero in the post-measurement states but they pick up different signs. After an average over all the measurement outcomes, we obtain an overall vanishing $\langle Z\rangle_{N_{\text{RG}}}$. In other words, the RG-assisted disorder parameter computed using a specific decoder is only reliable in the region where the 1-form symmetry is detected to exist by the decoder.

To further verify the nature of the transition, a finite-size data collapose is performed (see the inset in Fig.~\ref{Fig:app:product}b),
where we use the correlation length critical exponent $\nu=1.49(2)$ of the 2D RBIM at zero temperature~\cite{Preskill_2003}. Up to the reachable system sizes, the finite-size scaling is consistent with the universality of the 2D RBIM at zero temperature.

It will also be interesting to verify this transition by directly extracting the average of the topological entanglement entropy over sampled post-measurement states.

\section{Numerical algorithm for detecting emergent 1-form symmetries}\label{app:simulation_method}

Away from the product-state limit, we have to use numerical simulations to study the emergence of 1-form symmetries. 
Suppose we want to evaluate the expectation value of an operator $O$, e.g., a non-contractible $Z$ loop operator or a disorder operator from the QEC recovered states:   $\prod_{e\in \mathcal{S}_{\text{QEC}}(\pmb{m})}X_e\prod_p(1+m_pB_p)/2\ket{\Psi}$, where $\mathcal{S}_{\text{QEC}}(\pmb{m})$ is a recovery string configuration predicted by the MWPM decoder. The expectation value can be expressed as
\begin{align}\label{eq:expec_after_QEC}   
&\langle O\rangle=\notag\\
&\sum_{\pmb{m}}\tr\left[O\prod_{e\in \mathcal{S}_{\text{QEC}}(\pmb{m})}X_e\frac{1+m_pB_p}{2}\ket{\Psi}\bra{\Psi}\frac{1+m_pB_p}{2}\prod_{e\in \mathcal{S}_{\text{QEC}}(\pmb{m})}X_e\right]\notag\\
&=\sum_{\pmb{m}}\sum_{f(\pmb{s})={\pmb{m}}}\bra{\pmb{s}}\prod_{e\in \mathcal{S}_{\text{QEC}}(\pmb{m})}X_eO\prod_{e\in \mathcal{S}_{\text{QEC}}(\pmb{m})}X_e\ket{\pmb{s}}\bra{\pmb{s}}\ket{\Psi}\bra{\Psi}\ket{\pmb{s}},
\end{align}
where $\pmb{s}$ is a bit-string configuration, which is chosen in the $Z$ basis such that $\bra{\pmb{s}}\prod_{e\in \mathcal{S}_{\text{QEC}}(\pmb{m})}X_eO\prod_{e\in \mathcal{S}_{\text{QEC}}(\pmb{m})}X_e\ket{\pmb{s'}}$ is diagonal and can be easily evaluated on a classical computer, $f(\pmb{s})=\pmb{m}$ describes a bit string configuration such that $B_p\ket{\pmb{s}}=m_p\ket{\pmb{s}},\forall p$.
The only part that is not easy to calculate is the probability of the bit-string configuration $\pmb{s}$, given by $|\braket{\pmb{s}}{\Psi}|^2$, which we reformulate as a 2D tensor network.

\begin{figure}
    \centering
    \includegraphics{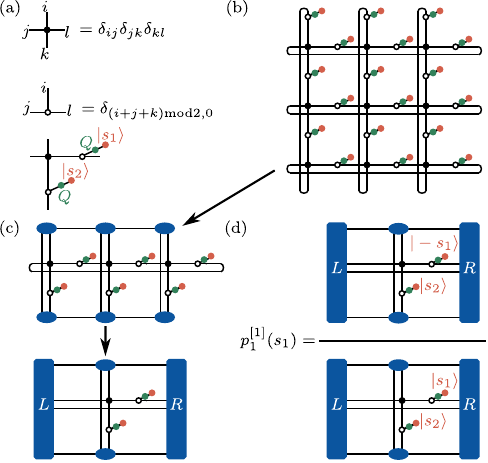}
    \caption{\textbf{Numerical algorithm for detecting emergent 1-form symmetries.} (a) The definition of the PEPS tensor for the deformed toric code states, in which the physical degrees of freedom are fixed after measurement. (b) The tensor network of $\braket{\pmb{s}}{\Psi}$ for periodic boundary conditions. (c) The 2D tensor network in (b) can be compressed to a 1D tensor network using the boundary MPS method, and the 1D tensor network is exactly contracted using the left and right boundary vectors $L$ and $R$.  (d) The relative probability for the Metropolis-Hasting algorithm is calculated using the environment of the tensor containing the flipped qubit.}
    \label{Fig:app_TNS}
\end{figure}

To this end, we combine tensor network methods and Monte Carlo sampling to evaluate Eq.~\eqref{eq:expec_after_QEC}. The method has been used to study physical problems related to RBIM~\cite{zhu_2023,Youjin_2024}. When $\ket{\Psi}$ is expressed in terms of a tensor network state on a torus, $\braket{\pmb{s}}{\Psi}$ is a tensor network shown in Fig.~\ref{Fig:app_TNS}. We use the Metropolis-Hastings
algorithm to sample bit strings, and the probability $|\braket{\pmb{s}}{\Psi}|^2$ is calculated using tensor networks. In the Metropolis-Hastings algorithm, we only need the relative probability between two bit-strings distinct by flipping one qubit:
\begin{equation}\label{eq:relative_prob}
p^{[1]}_e(\pmb{s})=\left|\frac{\bra{\pmb{s}}X_e\ket{\Psi}}{\braket{\pmb{s}}{\Psi}}\right|^2.
\end{equation}
This probability can be computed by contracting the tensor network on a torus. We double the virtual bond dimensions such that the tensor network on a torus in Fig.~\ref{Fig:app_TNS}b corresponds to a tensor network on a plane, which can be contracted using a finite boundary matrix product state (MPS) compression method; Fig.~\ref{Fig:app_TNS}c. During the MPS compression, we truncate the MPS bond dimension by discarding levels in the entanglement spectrum of the MPS that are smaller than $10^{-10}$, so that the effect of the finite bond dimension truncation is negiligible. After the MPS compression, the 2D tensor network is reduced to a 1D tensor network, which is exactly contracted using the boundary vectors $L$ and $R$, see Fig.~\ref{Fig:app_TNS}c. 
The relative probability in Eq.~\eqref{eq:relative_prob} is then calculated by constructing the environment of the tensor containing the flipped qubit, as shown in Fig.~\ref{Fig:app_TNS}d. 

After evaluating the relative probability $p_e^{[1]}$ of a single qubit flip, we can sample the bit-strings using the Metropolis-Hastings algorithm.  
If $p_e^{[1]}\geq 1$, we accept the single qubit flip. Otherwise, we accept the single qubit flip with the probability $p_e^{[1]}$. We sequentially flip all qubits on the edges of the lattice and determine the acceptance using the Metropolis-Hastings algorithm, such that parts of the environment of the previous tensor can be re-used to update the environment of the current tensor and the computational cost is largely reduced (as we do not have to re-contract the entire tensor network). 

In the presence of the bare $Z$ 1-form symmetry, the single qubit flip breaks the bare $Z$ 1-form symmetry,  $p_e^{[1]}$ is always $0$ and a qubit flip will never be accepted. When $g_x$ is small, $p_e^{[1]}$ is very small, and it is hard to accept a single qubit flip, so the sampling becomes inefficient. This problem can be circumvented by flipping a cluster of 4 qubits of a vertex and calculating the corresponding relative probability 
\begin{equation}\label{eq:relative_prob_4_flip}
p^{[4]}_e(\pmb{s})=\left|\frac{\bra{\pmb{s}}A_v\ket{\Psi}}{\braket{\pmb{s}}{\Psi}}\right|^2.
\end{equation}
We use the Metropolis-Hastings algorithm to decide whether we accept such a 4-qubit flip and sequentially sweep the 4-qubit flip for all vertices. In general, we can perform both single qubit flip and 4-qubit flips. 
With that we evaluate Eq.~\eqref{eq:expec_after_QEC} with sufficiently many samples to achieve convergence.

For the deformed toric code states, the local tensors on the vertices of the lattice are
\begin{equation}
    T^{s_1s_2}_{ijkl}=\sum_{t_1t_2mn}Q_{s_1t_1}Q_{s_2t_2}\delta_{(t_1+k+m)\text{mod} 2,0}\delta_{(t_2+l+n)\text{mod}2,0}\delta_{i,j}\delta_{j,m}\delta_{m,n},
\end{equation}
where $Q=1+g_x X+g_z Z$, the two physical legs $s_1,s_2$ are fixed by measuring $Z$ operators, and $i,j,k,l$ are four virtual legs, see Fig.~\ref{Fig:app_TNS}a. Then we implement the proposed protocol to detect the emergent 1-form symmetry.

For the toric code model in a field in Eq.~\eqref{eq:TC_Hamiltonianwithfields}, we can use a variationally optimized infinite projected entangled pair states (iPEPS)~\cite{PEPS_optimization_corboz,PEPS_optimization_laurens_V} to approximate its ground states. We use the single-site unit cell iPEPS ansatz with two physical legs for each tensor (similar to the tensor in Fig.~\ref{Fig:app_TNS}a but without the specific structure), and the variational optimization is performed using the PEPS-torch library~\cite{peps-torch}.
 The finite PEPS on a torus are created using the iPEPS tensors. Then we implement the proposed protocol to detect the emergent 1-form symmetry on this finite PEPS. Notice that, the iPEPS with a bond dimension $D$ has a correlation length $\xi_D$, the finite PEPS created using the iPEPS tensor is a relatively good approximate ground state of the finite system with the size $N_x\times N_y$ when $N_x\geq \xi_D$ and $N_y\geq \xi_D$. So when the iPEPS correlation length is large, i.e., near a quantum critical point, and we can not further increase $N_x$ and $N_y$ due to the computational cost, the error of finite PEPS is large so that the curve is not smooth in Fig.~\ref{Fig:scan_main}. Additionally, when the iPEPS undergo a continuous phase transition, the results from the finite PEPS created from iPEPS tensor would not have an expect scaling behavior. Therefore, %a better way to simulate our protocol is to%Since we only quantitatively show that our protocol works for the toric code mode in a field, and our accuracy is limited by the quality of the approximate ground states, see Appendix~\ref{app:simulation_method}.
 a future direction is finding a better way to accurately determine the information-theoretic transitions of 1-form symmetries and their universality classes. The possible approaches are using a sign problem free QMC method, or directly calculating the ground state of a finite system using large scale finite PEPS. These directions are technically more challenging and we leave it for further study.

\section{The $Z$ 1-form symmetry in the deformed toric code states along the $g_z=0$}\label{app:mapping_g_z=0}
In this section, we show that the information-theoretic transition of the $Z$ 1-form symmetry defined by Eq.~\eqref{eq:decodable} and the topological phase transition coincide along the line $g_z=0$. Consider the deformed toric code state along this line. The unnormalized wavefunction is
\begin{equation}
      \ket{\psi(g_x,0)}=\prod_e(1+g_x X_e)\ket{\tTC}.
\end{equation}
The normalization of the wavefunction is
\begin{equation}
    \bra{\text{TC}}\prod_e\prod_{e'}(1+g_x X_e)(1+g_xX_{e'})\ket{\tTC}.
\end{equation}
Observe that the expansion of the product is a linear combination of Pauli-$X$ string operators. Among the terms, only those that form closed contractible loops will evaluate to non-zero under $\ket{\tTC}$. By viewing the closed loops as domain walls of an Ising model on the dual lattice, (similar to Eq.~\eqref{app:eq:domain}) the expectation values of each Pauli-$X$ string operator can be mapped to the corresponding Boltzmann weights. For instance, we have the weights
\begin{equation}
    \includegraphics[scale =1]{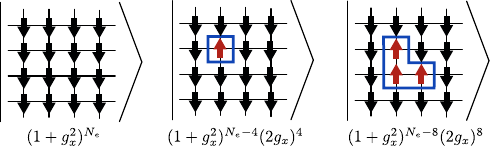}.
\end{equation}
Rewriting all the weights we get
\begin{align}
     \bra{\text{TC}}\prod_e\prod_{e'}&(1+g_x X_e)(1+g_xX_{e'})\ket{\tTC} 
     \nonumber \\
     &= \frac{[2g_x(1+g_x^2)]^{N_e/2}}{2}\mathcal{Z}_{\text{Ising}}(\tilde{J}),
\end{align}
where $\mathcal{Z}_{\text{Ising}}(\tilde{J}) = \sum_{\{\sigma_v\}}\exp\left[ \tilde{J}\sum_{\langle vv'\rangle}\sigma_v\sigma_{v'}\right]$ and $\exp(-2\tilde{J}) = 2g_x/(1+g_x^2)$. At the critical point of the 2D classical Ising model, the partition function, and therefore the wavefunction, exhibits critical behavior in their correlations. The transition from the topologically ordered to the trivial phases thus locates at the critical $\tilde{J}_c$ of the 2D classical Ising model. We proceed to compute the unnormalized probability weight of the 1-form charge configurations:
\begin{align}
    P_{\pmb{m}}=&\bra{\psi(g_x,0)}\prod_p\frac{1+m_pB_p}{2}\ket{\psi(g_x,0)}\notag\\
    =&\sum_{\pmb{q}}\left(\sum_{\mathcal{S}_{\pmb{m},\pmb{q}}}g_x^{|\mathcal{S}_{\pmb{m},\pmb{q}}|}\right)^2.
\end{align}
Using the trick in Appendix~\ref{app:map_to_RBIM}, we have 
\begin{align}\label{eq:sum_string_2_RBIM}
    \sum_{\mathcal{S}_{\pmb{m}, \pmb{q}}}g_x^{|\mathcal{S}_{\pmb{m}, \pmb{q}}|}=\frac{g_x^{N_e/2}}{2}\mathcal{Z}_{\text{RBIM}}(J,\mathcal{S}_{\pmb{m},\pmb{q}}),
\end{align}
where $J=-\log(g_x)/2$.
Using that the wavefunction normalization is $\bra{\text{TC}}\prod_e\prod_{e'}(1+g_x X_e)(1+g_xX_{e'})\ket{\tTC} =\sum_{\pmb{m}}P_{\pmb{m}}$, we get
\begin{equation}\label{app:eq:square_rbim}
    \sum_{\pmb{q}}\sum_{\pmb{m}}[\mathcal{Z}_{\text{RBIM}}(J,\mathcal{S}_{\pmb{m},\pmb{q}})]^2= 2[2g_x^{-1}(1+g_x^2)]^{N_e/2}\mathcal{Z}_{\text{Ising}}(\tilde{J}).
\end{equation}
On the intuitive level, squaring each $\mathcal{Z}_{\text{RBIM}}(J,\mathcal{S}_{\pmb{m},\pmb{q}})$ leads to terms which correspond to ``joining' endpoints of different string configurations deformable to $\mathcal{S}_{\pmb{m},\pmb{q}}$ to form closed domain walls, eventually resulting in the right hand side. Using similar reasoning, one can explicitly verify that~\cite{Sagar_ViJay_2024}
\begin{align}\label{app:eq:cross_rbim}
    \sum_{\pmb{q}}\sum_{\pmb{m}}\mathcal{Z}_{\text{RBIM}}(J,&\mathcal{S}_{\pmb{m},\pmb{q}})\mathcal{Z}_{\text{RBIM}}(J,\mathcal{S}_{\pmb{m},\pmb{q}}+\hat{\mathcal{C}}_x)\nonumber \\
   & = 2[2g_x^{-1}(1+g_x^2)]^{N_e/2}\mathcal{Z}^{\text{twist},x}_{\text{Ising}}(\tilde{J}),
\end{align}
where $\mathcal{Z}^{\text{twist},x}_{\text{Ising}}$ is the partition function of the 2D classical Ising model with twisted boundary condition in $x$ direction, i.e., the couplings along a non-contractible loop in $x$ direction are antiferromagnetic instead of ferromagnetic, and we can similarly define $\mathcal{Z}^{\text{twist},y}_{\text{Ising}}$  and $\mathcal{Z}^{\text{twist},x,y}_{\text{Ising}}$. The key quantity we want to examine is
\begin{equation}
     \lim_{N_x,N_y \rightarrow \infty} \sum_{\pmb{m}}\max_{\pmb{q}}\kappa_{\pmb{m},\pmb{q}} = \lim_{N_x,N_y \rightarrow \infty}  \frac{\sum_{\pmb{m}}\max_{\pmb{q} }\mathcal{Z}_{\text{RBIM}}(J,\mathcal{S}_{\pmb{m},\pmb{q}})^2}{\sum_{\pmb{m}',\pmb{q}'}\mathcal{Z}_{\text{RBIM}}(J,\mathcal{S}_{\pmb{m}',\pmb{q}'})^2}.
\end{equation}
In the ferromagnetic phase (the TC phase of $\ket{\psi(g_x,0}$), the domain walls cost extensive energy, we expect $\mathcal{Z}^{\text{twist},x}_{\text{Ising}}(\tilde{J})/\mathcal{Z}_{\text{Ising}}(\tilde{J}) = \mathcal{Z}^{\text{twist},y}_{\text{Ising}}(\tilde{J})/\mathcal{Z}_{\text{Ising}}(\tilde{J}) = \mathcal{Z}^{\text{twist},x,y}_{\text{Ising}}(\tilde{J})/\mathcal{Z}_{\text{Ising}}(\tilde{J})= \mathcal{Z}^{\text{twist},x,y}_{\text{Ising}}(\tilde{J})/\mathcal{Z}^{\text{twist},x}_{\text{Ising}}(\tilde{J})= \mathcal{Z}^{\text{twist},x,y}_{\text{Ising}}(\tilde{J})/\mathcal{Z}^{\text{twist},y}_{\text{Ising}}(\tilde{J}) = 0$ in the thermodynamic limit. Because all the partition functions are non-negative, Eqs.~\eqref{app:eq:square_rbim} and~\eqref{app:eq:cross_rbim} (and other variants associated with different twisted boundary conditions) imply that $ \lim_{N_x,N_y \rightarrow \infty} \sum_{\pmb{m}}\max_{\pmb{q}}\kappa_{\pmb{m},\pmb{q}} = 1$. More concretely, we note that for the $Z$ 1-form symmetry, there are four distinct $\pmb{q}$ for each given $\pmb{m}$. For notational simplicity, we denote the four $\mathcal{Z}_{\text{RBIM}}(J,\mathcal{S}_{\pmb{m},\pmb{q}})$ by $a_{\pmb{m}}, b_{\pmb{m}},c_{\pmb{m}}$ and $d_{\pmb{m}},$ such that $ a_{\pmb{m}}^2\geq  b_{\pmb{m}}^2\geq c_{\pmb{m}}^2\geq d_{\pmb{m}}^2$. Observe that $(a_{\pmb{m}}-d_{\pmb{m}}) - (b_{\pmb{m}}-c_{\pmb{m}})\leq a_{\pmb{m}}-d_{\pmb{m}}\leq a_{\pmb{m}}$, then
\begin{equation}\label{app:eq:lower_b}
   a_{\pmb{m}}^2\geq (a_{\pmb{m}} - d_{\pmb{m}} - b_{\pmb{m}} + c_{\pmb{m}} )^2.
\end{equation}
This implies
\begin{align}
   1\geq  &\lim_{N_x,N_y \rightarrow \infty} \frac{\sum_{\pmb{m}}a_{\pmb{m}}^2}{\sum_{\pmb{m}'}(a_{\pmb{m}'}^2+  b_{\pmb{m}'}^2+ c_{\pmb{m}'}^2+ d_{\pmb{m}'}^2)} \geq 
    \nonumber \\
    &\lim_{N_x,N_y \rightarrow \infty} \frac{\sum_{\pmb{m}}(a_{\pmb{m}}^2 +  b_{\pmb{m}}^2+ c_{\pmb{m}}^2+ d_{\pmb{m}}^2 + \text{ cross terms})}{\sum_{\pmb{m}'}(a_{\pmb{m}'}^2+  b_{\pmb{m}'}^2+ c_{\pmb{m}'}^2+ d_{\pmb{m}'}^2)} = 1,
\end{align}
where the cross terms come from the expansion of the right hand side of Eq.~\eqref{app:eq:lower_b}. The last equality follows from applying Eq.~\eqref{app:eq:cross_rbim} and its variants, knowing that the ratios between the partition functions with and without the twisted boundary condition are zero.

In the paramagnetic phase (the trivial phase of $\ket{\psi(g_x,0))}$, the domain walls condense, we expect $\mathcal{Z}^{\text{twist},x}_{\text{Ising}}(\tilde{J})/\mathcal{Z}_{\text{Ising}}(\tilde{J}) = \mathcal{Z}^{\text{twist},y}_{\text{Ising}}(\tilde{J})/\mathcal{Z}_{\text{Ising}}(\tilde{J}) = \mathcal{Z}^{\text{twist},x,y}_{\text{Ising}}(\tilde{J})/\mathcal{Z}_{\text{Ising}}(\tilde{J}) = 1$ in the thermodynamic limit. Eqs.~\eqref{app:eq:square_rbim} and~\eqref{app:eq:cross_rbim} (and other variants) imply that, for a random $\pmb{m}$ and $\mathcal{N}_{\pmb{m}} =a_{\pmb{m}}^2 +  b_{\pmb{m}}^2+ c_{\pmb{m}}^2+ d_{\pmb{m}}^2$, we have $a_{\pmb{m}}^2/\mathcal{N_{\pmb{m}}} = b_{\pmb{m}}^2/\mathcal{N_{\pmb{m}}}  = c_{\pmb{m}}^2/\mathcal{N_{\pmb{m}}}  = d_{\pmb{m}}^2/\mathcal{N_{\pmb{m}}} $ with probability one, we therefore have $ \lim_{N_x,N_y \rightarrow \infty} \sum_{\pmb{m}}\max_{\pmb{q}}\kappa_{\pmb{m},\pmb{q}} = 1/4$.

\section{Equivalence between the FM string order parameter and the disorder parameter measured from QEC recovered states}\label{EC_string_order_and_FM}
In the main text, we show that the disorder parameter can be used to detect the topological phase transition after successfully recovering the bare 1-form symmetry.
In this section, we show that the disorder parameter (without RG-assistance) measured on the QEC-recovered states and the FM string order parameter~\cite{FM_1983,FM_1986} are equivalent if we can successfully recover the bare 1-form symmetry. 

The FM string order parameter is defined as
 \begin{equation}\label{FM_op}
   O_{\text{FM}}(|\mathcal{C}_{1/2}|)=\frac{\langle{\Psi}|\prod_{e\in \mathcal{C}_{1/2}}Z_e|{\Psi}\rangle}{\sqrt{\bra{\Psi}\prod_{e\in \mathcal{C}} Z_e\ket{\Psi}}},   
 \end{equation}
 where $|\mathcal{C}_{1/2}|$ is the length of the string $\mathcal{C}_{1/2}$, $\mathcal{C}$ is a close loop with a length $|\mathcal{C}|=2|\mathcal{C}_{1/2}|$. When the $Z$ 1-form symmetry is an emergent symmetry, the numerator decays exponentially to zero with $|\mathcal{C}_{1/2}|$ in any phase, so the denominator is taken into consideration to cancel the exponential decay of the numerator, such that only the end points of the $Z$ string contribute to $O_{\text{FM}}$. In the presence of the $Z$ 1-form symmetry, the FM string order parameter is zero in the topological phase and non-zero in the trivial phase.

  \begin{figure}
    \centering
    \includegraphics[scale=0.5]{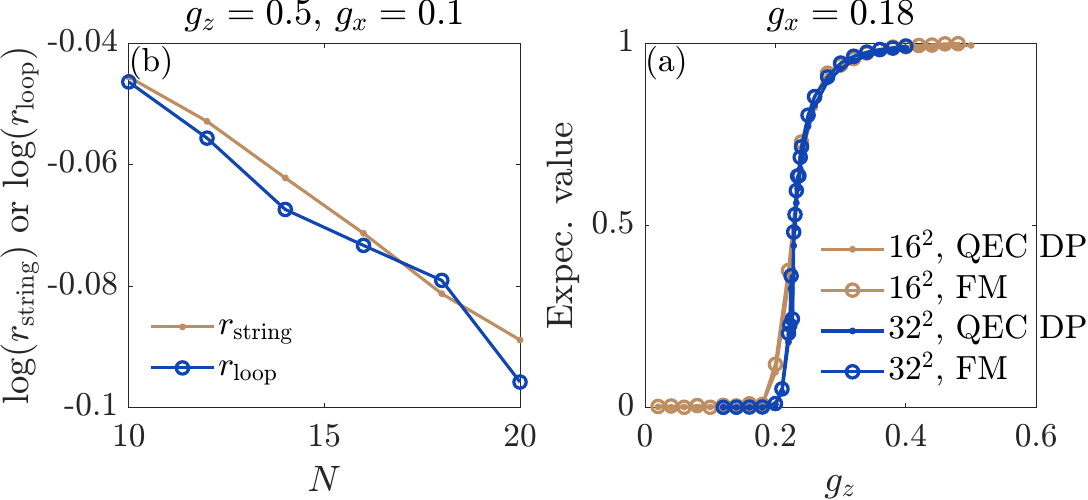}
    \caption{\textbf{Comparison of the FM string order parameter measured from the original state and the disorder parameter measured from the QEC-recovered states.} (a) The ratio defined in Eq.~\eqref{eq:ratio} calculated at a point $(g_x,g_z)=(0.1,0.5)$ from the deformed toric code states with different sizes $N\times N$, and the length of the string (loop) is $|\mathcal{C}_{1/2}|=N/2$ ($|\mathcal{C}|=N$). (b) FM string order parameter measured from the original quantum states and the disorder parameter measured from the QEC-recovered states (QEC DP) along $g_x=0.18$ of the deformed toric code states. We consider two different system size $16\times 16$ and $32\times 32$, where the length $|\mathcal{C}_{1/2}|$ of the string is $8$ and $16$, respectively, showing perfect agreement (dots and circles).   
     }
    \label{Fig:compare_ECS_and_FM}
\end{figure}

We now relate the FM string order parameter measured from the original state with the emergent $Z$ 1-form symmetry to the disorder parameter measured from the QEC-recovered states with the bare $Z$ 1-form symmetry. By defining a trace preserving quantum channel 
 \begin{equation}
    \mathscr{N}[\cdot]=\sum_{\pmb{m}}\tr\left[\prod_{e\in \mathcal{S}_{\text{QEC}}(\pmb{m})}X_e\frac{1+m_pB_p}{2}\cdot \frac{1+m_pB_p}{2}\prod_{e\in \mathcal{S}_{\text{QEC}}(\pmb{m})}X_e\right], 
 \end{equation}
 we have
 \begin{align}
      O_{\text{FM}}(|L_{1/2}|)&=\frac{\Tr\left(\prod_{e \in \mathcal{C}_{1/2}}Z_e|{\Psi}\rangle\langle{\Psi}|\right)}{\sqrt{\Tr\left(\prod_{e\in L} Z_e\ket{\Psi}\bra{\Psi}\right)}}\notag\\
      &=\frac{\Tr\left(\mathscr{N}\left[\prod_{e \in \mathcal{C}_{1/2}}Z_e|{\Psi}\rangle\langle{\Psi}|\right]\right)}{\sqrt{\Tr\mathscr{N}\left[\left(\prod_{e \in \mathcal{C}} Z_e\ket{\Psi}\bra{\Psi}]\right)\right]}}.\label{eq:step2}
 \end{align}
To relate to the disorder parameter measured on the QEC-recovered states, we need to take the $Z$ string (loop) operator in the numerator (denominator) out of the quantum channel $\mathscr{N}[\cdot]$. Because the recovery $X$ strings $\prod_{e\in \mathcal{S}_{\text{QEC}}(\pmb{m})}X_e$ and the $Z$ string (loop) anti-commute with each other at their crossing points, we suppose that in a unit length of the $Z$ string, the probability of even and odd number of crossings is $p$ and $1-p$, separately. Then for an even length $L$ the probability of even (odd) number of crossings is $\sum_{m=0}^{L/2}C_{L}^{2m} p^{2m}(1-p)^{{L}-2m}$  [$\sum_{m=0}^{L/2}C_{L}^{2m-1} p^{2m-1}(1-p)^{L-2m+1}$ ], where $C_{L}^{m}$ is the binomial. So by taking the $Z$ string (loop) out of the quantum channel we expect this operation will result in an exponential decay coefficient, namely 
 \begin{align}
    r_{\text{string}}=\frac{\Tr\{\mathscr{N}[\prod_{e \in \mathcal{C}_{1/2}} Z_e\ket{\Psi}\bra{\Psi}]\}}{\Tr\{\prod_{e \in \mathcal{C}_{1/2}} Z_e\mathscr{N}[\ket{\Psi}\bra{\Psi}]\}}\sim (2p-1)^{|\mathcal{C}_{1/2}|} ,\notag\\
    r_{\text{loop}}=\sqrt{\frac{\Tr\{\mathscr{N}[\prod_{e \in \mathcal{C}} Z_e\ket{\Psi}\bra{\Psi}]\}}{\Tr\{\prod_{e \in \mathcal{C}} Z_e\mathscr{N}[\ket{\Psi}\bra{\Psi}]\}}}\sim (2p-1)^{|\mathcal{C}|/2}. \label{eq:ratio}
 \end{align}
We also numerically check Eq.~\eqref{eq:ratio} in Fig.~\ref{Fig:compare_ECS_and_FM}a, which indicates that both the ratios decays exponentially with the same rate. So we can further simplify Eq.~\eqref{eq:step2}:
 \begin{align}
    &O_{\text{FM}}\approx\frac{ e^{-\alpha |\mathcal{C}_{1/2}|}\Tr\left( \prod_{e \in \mathcal{C}_{1/2}}Z_e \mathscr{N}\left[|{\Psi}\rangle\langle{\Psi}|\right]\right)}{ \sqrt{e^{-\alpha |\mathcal{C}|}\Tr\left\{\prod_{e \in \mathcal{C}} Z_e\mathscr{N}\left[\left(\ket{\Psi}\bra{\Psi}]\right)\right]\right\}}}\label{eq:step3}\\
       &=\Tr\left( \prod_{e \in \mathcal{C}_{1/2}}Z_e \mathscr{N}\left[|{\Psi}\rangle\langle{\Psi}|\right]\right).
 \end{align}
 We conclude that the FM string order parameter measured from the original quantum state and the disorder parameter measured from the QEC-recovered states are the same when $|\mathcal{C}_{1/2}|\rightarrow \infty$, 
 and the numerical comparison between the FM string order parameter and the disorder parameter from the QEC-recovered states shown in Fig.~\ref{Fig:compare_ECS_and_FM}b supports the conclusion.   

\section{Renormalization group algorithm}\label{app:QCNN}

\begin{figure}
    \centering
    \includegraphics[width=7.5cm]{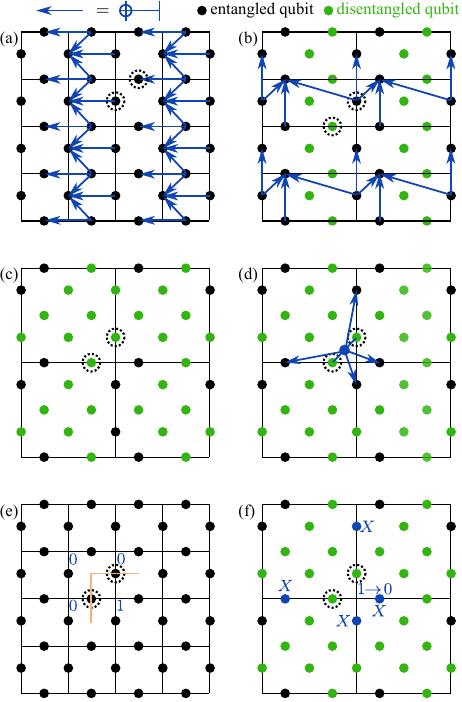}
    \caption{\textbf{The steps of the classical RG algorithm.} We represent the CNOT gates using arrows for simplicity. The green and black dots denote the disentangled qubits and entangled qubits, respectively. The qubits in the dashed-line circles are the control qubits whose values in the $Z$ basis are conserved. The boundary condition of the lattice is periodic. (a) The horizontal coarse graining step. (b) The vertical coarse graining step. (c) The coarse-grained lattice. (d) The local operation to keep the domain wall non-growing, which is an control-$A_v$ operator applying on the vertex of the coarse-grained lattice and controlled by the control qubits. (e) Before the coarse graining step, when both control qubits are 1, it indicates a corner of the domain wall. A possible Ising spin configuration corresponding to the domain wall is shown, where $0$ ($1$) represents spin-up (spin-down). (f) If both control qubits are $1$, we try to shrink the domain wall by applying an $A_v$ operator on the coarse-grained lattice.}
    \label{Fig:app_QCNN}
\end{figure}
In this section, we describe the classical RG algorithm for assisting the measurement of the disorder parameter. 
The algorithm is applied to a classical qubit (spin) configuration sampled from the wavefunction by $Z$
-basis measurements. Each step of the RG contains two parts, the coarse graining of the lattice and local processing to shrink the loops (domain walls). The coarse graining is performed using CNOT gates which disentangle some degrees of freedom from the original lattice~\cite{chu_2023_RG} (A CNOT gate applies a Pauli $X$ to the target qubit when the control qubit is $\ket{1}$ and otherwise it acts trivially). As shown in Fig.~\ref{Fig:app_QCNN}a and b, we first apply the CNOT gates horizontally, the size of a system reduces from $N_v\times N_h$ to $N_v\times (N_h/2)$, where $N_v$ and $N_h$ are the linear system sizes in the vertical and horizontal directions. Then we apply the CNOT gates vertically, the system size further reduces from $N_v\times (N_h/2)$ to $(N_v/2)\times (N_h/2)$, see Fig.~\ref{Fig:app_QCNN}c.  Fig.~\ref{Fig:app_QCNN}a-c count as one step of the lattice coarse graining. 

In the presence of the bare $Z$ 1-form symmetry, the symmetric subspace is spanned by closed loop configurations formed by strings of $\ket{1}$. It can be verified that the closed loop constraint is preserved on the coarse-grained new lattice.
After the coarse graining, we perform a local operation to shrink the size of the loops. The idea is that the loops perpendicular to the edges can be regarded as the domain walls of the (virtual) Ising spins at the vertices, a local operation based on the majority vote of the Ising spins is then used to shrink the domain walls. As shown in Fig.~\ref{Fig:app_QCNN}d, the local processing has the following explicit rule: if both of the two qubits within the vertex of the new lattice are $1$, then we apply a vertex operator $A_v$ to the vertex of the coarse-grained lattice, otherwise we do nothing. This is achieved by a control-$A_v$ gate with two control qubits. The local rule is designed based on the observation that the states of the two control qubits in the $Z$ basis are not changed by the lattice coarse-graining circuit; when both two control qubits are 1, they correspond to a corner of the domain wall, as shown in Fig.~\ref{Fig:app_QCNN}e and f. To keep the size of the domain walls non-increasing, we apply an $A_v$ to the coarse-grained lattice, which effectively flips the Ising spin at the vertex of the coarse-grained lattice.  
After the virtual spin flip, the total number of domain walls may either remain the same (e.g., if we are at the corner of a very large domain) or decrease (e.g., if we are at the corner of a domain wall surrounding a single plaquette). Although both scenarios are possible, repeating this process will eventually drive the total number of domain walls down to zero in the trivial phase. This is because, most of the time, the number of domain walls stays the same, and occasionally it shrinks, so the algorithm’s fixed point ends up being either a state with no domain walls or a completely disordered arrangement of domain walls.

After the coarse graining and the local processing, one step of the RG is finished. We can perform $(N-1)$ steps of the RG on a system with a size $N_x=N_y=2^N$ such that the system size reduces to $2\times 2$. After each RG step, we measure a single $Z$ operator, which effectively corresponds to a $Z$ string with a length $2^{N_{\text{RG}}}$ on the original lattice, where $N_{\text{RG}}$ is the number of the performed RG steps. If the expectation value of the single $Z$ operator approaches $1$ as $N_{\text{RG}}$ increases, the quantum state is topologically trivial. If the quantum state is topologically ordered (or more precisely, the 1-form symmetry is spontaneously broken), the expectation value of the single $Z$ operator approaches $0$ as $N_{\text{RG}}$ increases.

As a side remark, the RG algorithm can alternatively be executed as a unitary quantum circuit on the QEC-recovered quantum state, which provides a concrete example of a 2D quantum convolutional neural network (QCNN) previously investigated for quantum phase recognition in 1D quantum systems~\cite{QCNN_2019,RG_QEC_exact_2023,QCNN_Liu_2023}. The unitary protocol maintains the coherence of the state during the coarse graining, allowing for the potential use of the coarse grained states as useful quantum resource.

\begin{figure*}
    \centering
    \includegraphics[scale=0.5]{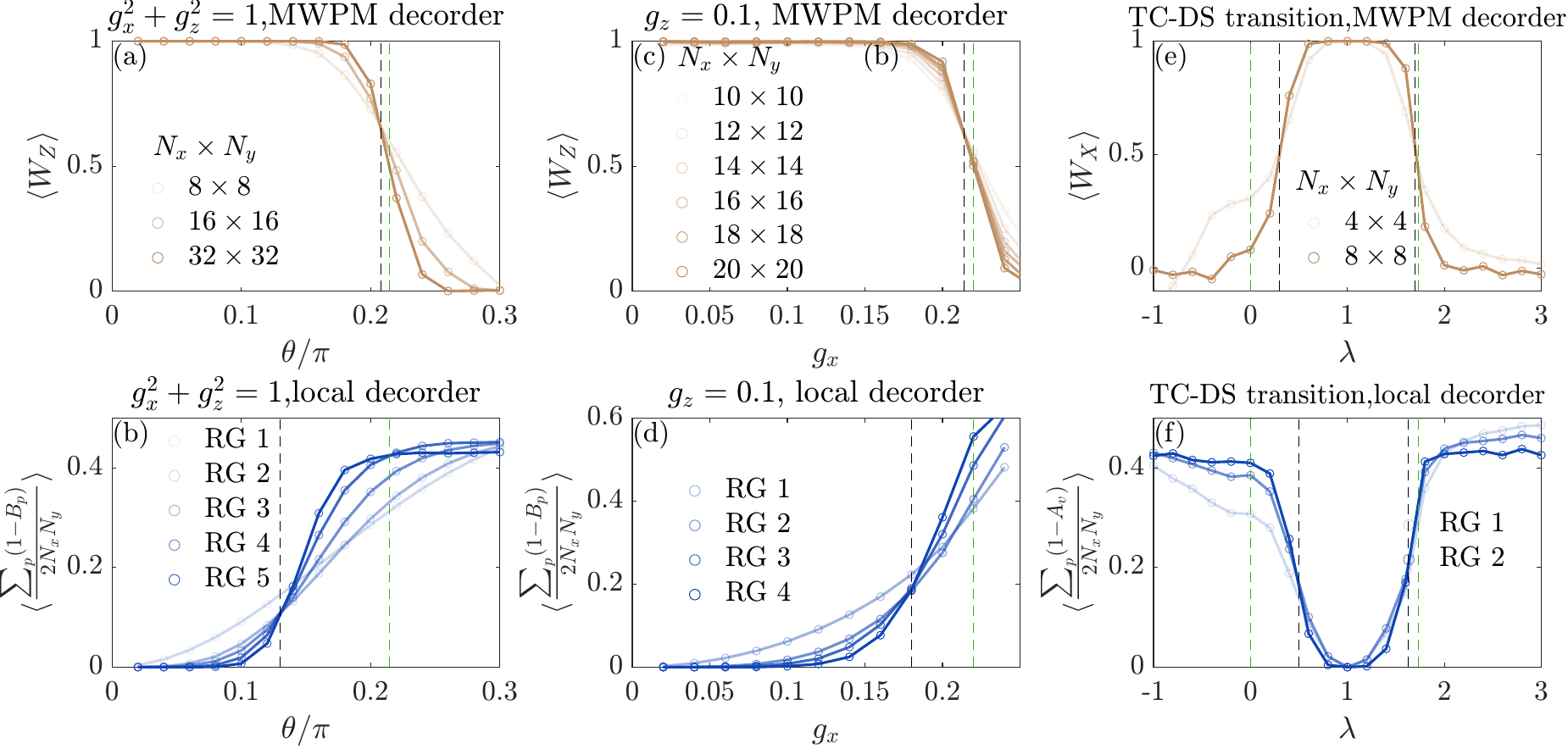}
    \caption{\textbf{Comparison of the thresholds given by the global decoder and the local decoder.} The green dashed lines are the optimal thresholds, the black dashed lines are the thresholds predicted by the specific decoders we use. The subplots in the first row are expectation values $\langle W_Z\rangle$ or $\langle W_X\rangle$ of the non-contractible loop operators measured from the MWPM decoder recovered states. $N_x\times N_y$ is the system size. The subplots in the second row are the 1-form symmetry charge densities $\langle\sum_p(1-B_p)/(2N_xN_y)\rangle$ or $\langle\sum_v(1-A_v)/(2N_xN_y)\rangle$ evaluated after performing the local QEC shown in Eq.~\eqref{Eq:local_decorder}.  First column: product states. Middle column:  deformed toric code states of Eq.~\eqref{eq:deformed_wavefunction} along $g_z=0.1$ line. Right column: deformed double-semion quantum states defined by Figs.~\ref{Fig:TC_DS}a and b as well as Eq.~\eqref{eq:def_double_line_tensor}. 
     }
    \label{Fig:compare_local_global_decorder}
\end{figure*}

\section{The performance of the global MWPM decoder and a local decoder}\label{app:MWPM_vs_local_decorder}

The idea of detecting the emergent higher-form symmetries using QEC can be performed using any decoder. Each of them can have a different performance, as characterized by the distance between threshold determined by the decoder and the optimal threshold. Previous work has applied various local decoders to detect topological order~\cite{LED_2024}. 
In this appendix, we compare a local decoder defined in Ref.~\cite{Timothy_2024} and the global MWPM decoder used in the main text and show that the global QEC decoder performs much better than the local QEC decoder.  

Let us describe the definition of the specific local decoder we implement. We can define $2\times 2$ blocks in each of the two sublattices and they are marked using the blue and orange colors:   
\begin{equation}\label{Eq:local_decorder}
\vcenter{\hbox{
  \includegraphics[scale=0.5]{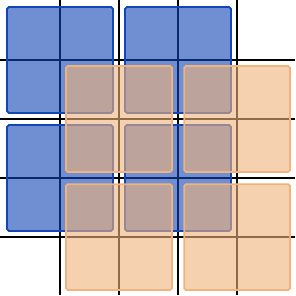}}}.
\end{equation}
At first we measure all $B_p$ operators to get the syndrome configurations. Then, in each $2\times 2$ blue block in Eq.~\eqref{Eq:local_decorder}, if there are an even number of $Z$ 1-form symmetry charges, we can directly remove all of them; if there are an odd number of $Z$ 1-form symmetry charges, we do nothing. Next, we process to next step by looking at each orange $2\times 2$ block in Eq.~\eqref{Eq:local_decorder}, if there are an even number of $Z$ 1-form symmetry charges, we can directly remove them; if there are an odd number of $Z$ 1-form symmetry charges, we fuse them to one charge and move it to the left-upper corner of the orange block. Finally, we can renormalize the lattice such that the orange $2\times 2$ blocks become a $1\times 1$ plaquette of the new lattice. We then repeatedly perform the procedures. After each renormalization, we  measure the $Z$ 1-form symmetry charge density, which is defined as $\sum_p(1-B_p)/(2N_xN_y)$, where $N_x\times N_y$ is the size of the renormalized lattice. 
The success of the decoder is marked by the ability to remove all the 1-form symmetry charges from the lattice, i.e., the 1-form symmetry charge density approaches 0 with more RG steps.

In Fig.~\ref{Fig:compare_local_global_decorder}, we compare the thresholds obtained from the local decoder and the global MWPM decoder using three examples: i) the 2D product states; ii) the deformed toric code states in Eq.~\eqref{eq:deformed_wavefunction} with $g_z=0.1$; iii) the quantum states describing the phase transition between the TC phase and the double-semion phase. It can be found that the thresholds determined by the global MWPM decoder are significantly closer to the optimal thresholds than those determined by the local decoder in all examples. 
We expect that generally, the performance of a well-chosen global decoder is better than the local decoders as they allow non-local processing of the classical information.

% \begin{figure}
%     \centering
%     \includegraphics[scale=0.5]{Fig_app_subsystem.pdf}
%     \caption{} 
%     \label{Fig:result_sub_sys_protocol_SPT}
% \end{figure}

\section{Applying the protocol on a deformed toric code}
\label{app:sec:deformTC}
\begin{figure}
    \centering
    \includegraphics[scale=0.5]{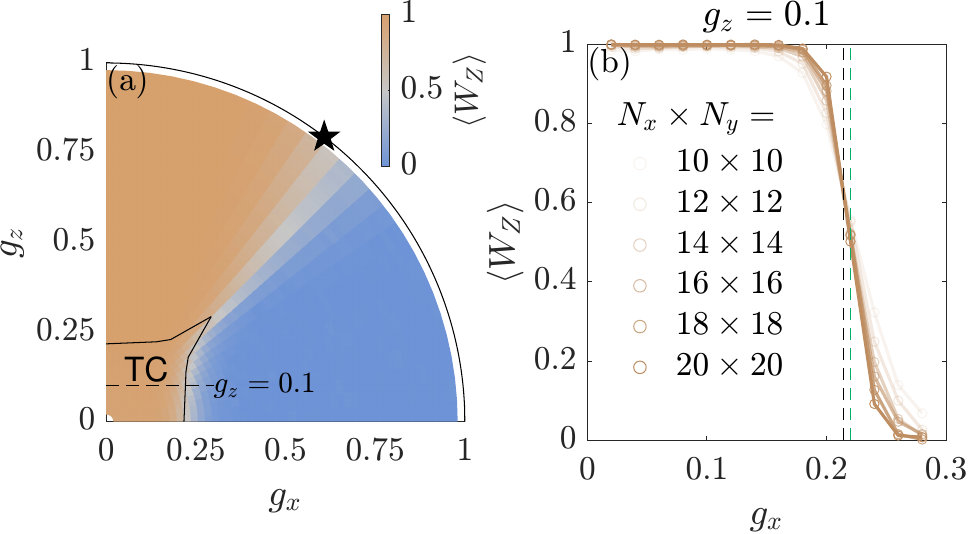}
    \caption{\textbf{Numerical results for detecting $Z$ 1-form symmetry from the deformed toric code states.} 
   (a) The colormap for the expectation values of the non-contractible $Z$ loop operator $\langle W_Z\rangle$ from the QEC-recovered states. A value of +1 indicates the existence of the 1-form symmetry. The star marks the exact field where the emergent Z 1-form symmetry ceases to exist in the product-state limit. We scan the entire parameter space with a system of $N_x \times N_y = 8\times 8$ plaquettes. Cuts of the phase diagram (b) for $g_z=0.1$. The black and green dashed lines mark the numerical and the theoretical transition points, respectively.}
    \label{Fig:scan}
\end{figure}

% \begin{figure}
%     \centering
%     \includegraphics[scale=0.5]{topological_transition.pdf}
%     \caption{\textbf{Detecting topological phase transition using the disorder parameter from the QEC-recovered deformed toric states.} (a) The colormap for the expectation values of the RG-assisted disorder parameter $\langle Z\rangle_{{N}_{\text{RG}}}$ with $N_{\text{RG}}=2$ on the QEC-recovered state. The expectation is computed using a single $Z$ operator evaluated after applying 2 steps of the classical RG processing. The detection protocol is applied to the region with the $Z$ 1-form symmetry (see the orange region in Fig.~\ref{Fig:scan}b).
%      (b) Cut along the line $g_z=0.18$ shown in (a). 
%      The inset in (b) shows the data collapse using the correlation length critical exponent $\nu=1$ of the 2D classical Ising model, and $g_{z,c}\approx 0.229$. }
%     \label{Fig:inf_field}
% \end{figure}

In this section we show detailed numerical results of the $Z$ 1-form symmetry indicator evaluated from the deformed toric code states in Eq.~\eqref{eq:deformed_wavefunction}. The 1-form symmetry and the phase diagram for the deformed toric code state are shown in Fig.~\ref{fig:deformTC}a. %The phase boundary between the TC phase and the trivial phase is obtained by mapping the deformed toric code states in Eq.~\eqref{eq:deformed_wavefunction} to a partition function of a 2D classical statistical mechanics model~\cite{Zhu_2019}. The information-theoretic transition of the 1-form symmetries is independent of the quantum phases. 

We apply the protocol to numerically scan the information-theoretic transition of the $Z$ 1-form symmetry in the phase diagram of the deformed toric code states. 
The $Z$ 1-form symmetry indictor $\langle W_Z\rangle$ measured from the QEC recovered states is shown in Fig.~\ref{Fig:scan}a, which implies in the toric code phase and a part of the trivial phase have the $Z$ 1-form symmetry. In Fig.~\ref{fig:deformTC}a, and we evaluate the RG-assistant disorder parameter in the orange region of Fig.~\ref{Fig:scan}a.

We now benchmark MWPM decoder in the vicinity of the topological phase transition. Concretely, we focus on the line $g_z = 0.1$ in Fig.~\ref{Fig:scan}b. %The numerically obtained transition point from the MWPM decoder is again slightly below the critical point for the topological phase transition due to the sub-optimality,  see Fig.~\ref{Fig:scan}b. 
The points of the transition in Fig.~\ref{Fig:scan}b are obtained from the MWPM decoder, which is not optimal and underestimates the correct threshold upon which the 1-form symmetry is absent. For this reason, it will not coincide with the phase boundary (blue lines in Fig.~\ref{fig:deformTC}a) between the TC phase and the trivial phase, but approximates it reasonably well.
The information-theoretic transition boundary of the $Z$ 1-form symmetry qualitatively agrees with the theoretical phase diagram Fig.~\ref{fig:deformTC}a. 

%We next numerically evaluate cuts in the product-state limit $g_x^2+g_z^2 = 1$ using the MWPM decoder. As discussed in Sec.~\ref{sec:2dproduct}, the $Z$ 1-form symmetry exists when $\theta\in[0,\theta_c)$, where $\theta_c= 0.22146(2)\pi$. 
%If an optimal decoder is used, the predicted transition corresponds to the RBIM along the Nishimori line~\cite{Nishimori_1981}. However, the MWPM decoder is not optimal and predicts a lower threshold than the optimal decoder. Interestingly, MWPM decoder corresponds to the RBIM at zero temperature~\cite{Preskill_2003}. It follows that the MWPM decoder correctly identifies the 1-form symmetry up to $\theta_c^{\text{MWPM}}= 0.2081(1)\pi < \theta_c$, consistent with our $Z$ 1-form symmetry indicator $\langle W_Z\rangle$ shown in Fig.~\ref{Fig:scan}c. As the system size increases from $8\times 8$ to $32\times 32$ plaquettes, the indicator converges to +1 for $\theta<\theta_c^{\text{MWPM}}$ and to 0 for $\theta > \theta_c^{\text{MWPM}}$. Nonetheless, the MWPM decoder is a reasonably good  decoder in practice for detecting the 1-form symmetries, with a predicted threshold very close to the theoretically optimal threshold. 

\section{Applying the protocol on transition between TC and the double-semion phase}\label{sec:tctods}

Ground states with distinct topological order can be characterized by different 1-form symmetries. In such cases the detection of an emergent 1-form symmetry can be used to get useful information about the topological phases as well. As an example, we consider the double-semion model~\cite{Freedman:2004,Levin_Wen_2005} which has $\mathbb{Z}_2$ topological order twisted by non-trivial 3-cocycles~\cite{Dijkgraaf_Witten_1990,Levin_Gu_2012}. The TC phase and the double-semion phase have the same topological entanglement entropy~\cite{TEE_levin_Wen_2006,TEE_Kitaev_Preskill_2006}, making them hard to distinguish from this perspective. Nevertheless, we can use the QEC protocol to distinguish the TC phase from the double-semion phase using the fact that only the TC phase can have the $X$ 1-form symmetry coexisting with the $Z$ 1-form symmetry but the double-semion phase does not.  

\begin{figure}
    \centering
    \includegraphics[scale=0.4]{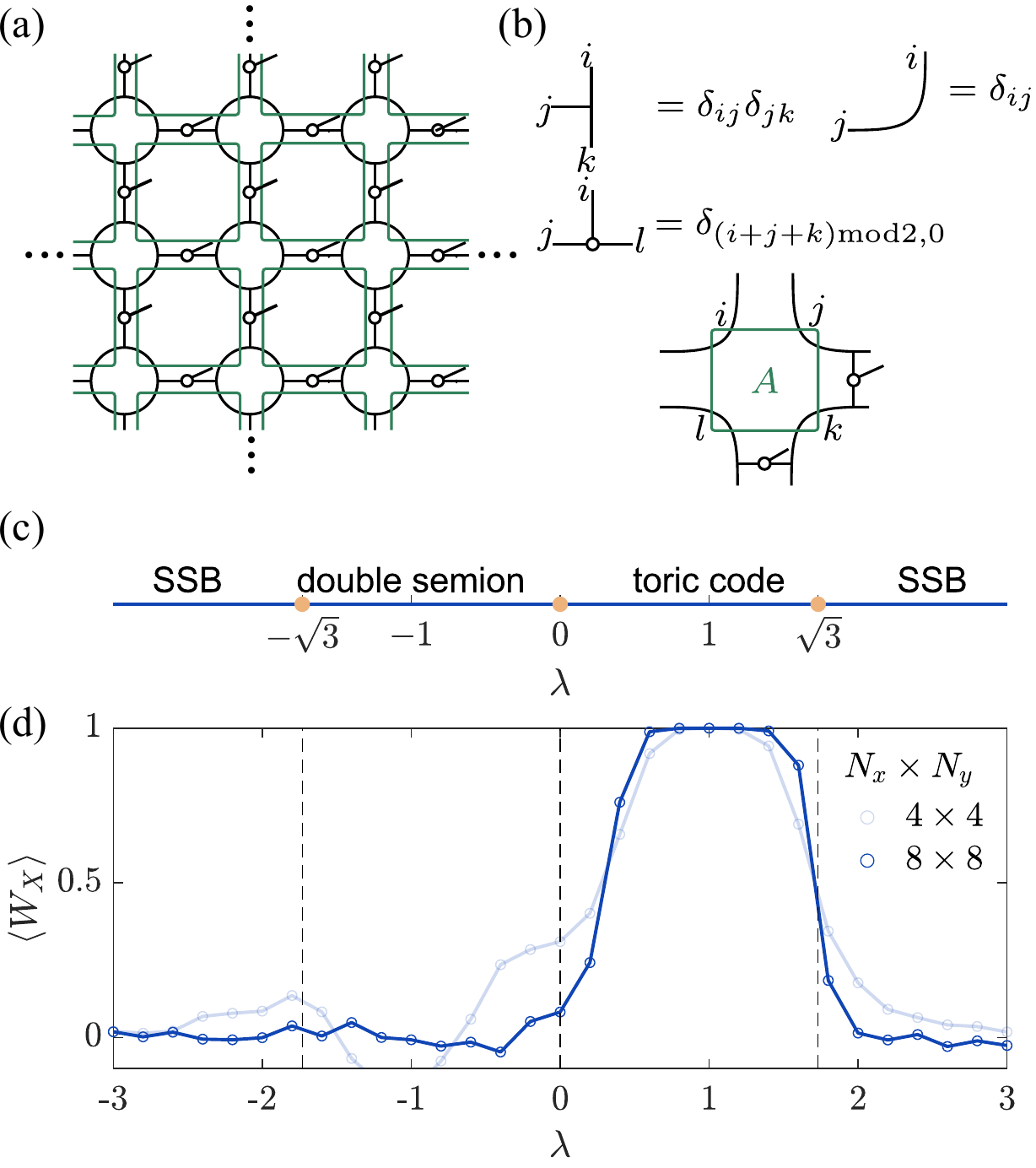}
    \caption{\textbf{Detecting the phase transition between the TC phase and double-semion phase via the $X$ 1-form symmetry.}  (a) The toy quantum states are given by the double-line tensor network states obtained by contacting virtual legs of the double line tensors. (b) Definition of the double-line tensor, where the two legs of the same line have to be the same so we label them using a single index. The tensor $A_{ijkl}$ is defined in Eq.~\eqref{eq:def_double_line_tensor}.  
    (c) The phase diagram, which contains the TC phase, the double-semion phase as well as the spontaneous symmetry breaking (SSB) phases which break the lattice rotation symmetry.   (d) Expectation value of a non-contractible $X$ loop measured from the QEC-recovered states approaches one in the TC phase but vanishes in the double-semion phase where it is not an emergent 1-form symmetry.}
    \label{Fig:TC_DS}
\end{figure}

We apply the protocol to a family of simple quantum states realizing a topological phase transition between the TC phase and the double-semion phase~\cite{Xu_2018}. The toy quantum states are constructed by gauging 2D $\mathbb{Z}_2$ symmetry protected topological states~\cite{Levin_Gu_2012,Xie_chen_SPT_2011,Huang_Wei_2016}. They are defined using the double-line tensor-network states (see Figs.~\ref{Fig:TC_DS}a and b), where the tensor $A$ has a tuning parameter $\lambda$:
\begin{equation}\label{eq:def_double_line_tensor}
    A_{0011}=A_{0110}=\lambda;\,\,\, A_{1100}=A_{1001}=|\lambda|;\,\,\, 
    A_{ijkl}=1, \mbox{otherwise}.
\end{equation}
The phase diagram of the tunable tensor-network states is shown in Fig.~\ref{Fig:TC_DS}c. In addition to the TC phase and the double-semion phase, the system also exhibits stripe order as a conventional spontaneous symmetry breaking phase, which breaks the lattice rotation symmetry~\cite{Xu_2018}. 

For all $\lambda$, the ground state has the bare $Z$ 1-form symmetry. The toric code ground state at $\lambda=1$ also has the bare $X$ 1-form symmetry. In the double-semion ground state at $\lambda=-1$, no $X$ 1-form symmetry exists but there is another 1-form symmetry generated by some loop operators~\cite{Levin_Wen_2005}. We apply the protocol to detect the $X$ 1-form symmetry.
The numerical results obtained from our protocol are shown in Fig.~\ref{Fig:TC_DS}d. In the toric code phase the expectation value of the 1-form symmetry indicator $\langle W_X\rangle$ approaches $1$ as the system size increases, where $W_X=W_X^{[x]}+W_X^{[y]}$, and $W_X^{[x]}$ and $W_X^{[y]}$ are non-contractible $X$ loop operators on the dual lattice. However, in the double-semion phase as well as in the symmetry broken phases, the emergent $X$ 1-form symmetry ceases to exist. The protocol can therefore be used to efficiently infer the existence of specific topological order which could otherwise be hard to resolve, for instance by means of the topological entanglement entropy.

We comment on the simulation of applying the protocol on the toy double-line tensor network states shown in Fig.~\ref{Fig:TC_DS}a. We pair the virtual legs of the double line tensor in Fig.~\ref{Fig:TC_DS}b such that it is reshaped to a tensor with virtual bond dimension 4. Since both the toric code model and the double-semion model have the $Z$ 1-form symmetry,  we aim for detecting the $X$ 1-form symmetry in order to distinguish them. In the simulation, we therefore use $X$ basis as the computational basis. Then we can proceed similarly as the case of the deformed TC states introduced above. 
In addition, since the quantum states in Fig.~\ref{Fig:TC_DS}a has the global $\mathbb{Z}_2$ symmetry $\prod_eX_e$, single qubit flip in the $X$ basis  will never be accepted in the Monte Carlo sampling, and we instead use two-qubit flips to sample. 

\section{Applying the protocol to the Ising chain}\label{sec:Ising_chain}
\begin{figure}
    \centering
    \includegraphics[width=1\linewidth]{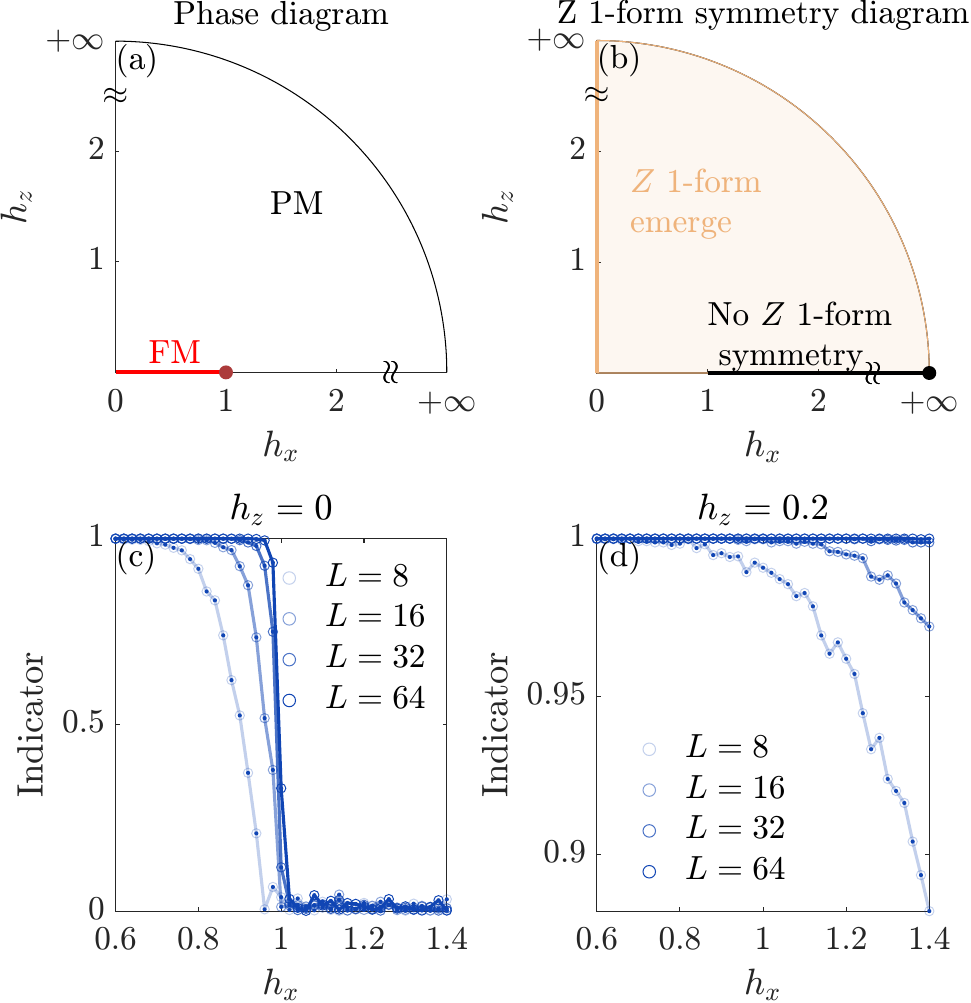}
    \caption{\textbf{Detecting the $Z$ 1-form symmetry in the quantum Ising chain.}  FM (PM) denotes the ferromagnetic (paramagnetic) phase. (a) The physical phase diagram. (b) The information-theoretic phase diagram of the $Z$ 1-form symmetry. The $Z$ 1-form symmetry is exact when  $h_x=0$. (c) $Z$ 1-form symmetry indicator along $h_z=0$. (c) $Z$ 1-form symmetry indicator along $h_z=0.2$. $L$ is the length of the periodic MPS. }
    \label{Fig:Ising_chain}
\end{figure}

Here we discuss another example of non-frustration-free Hamiltonians with degenerate ground states. We apply our protocol to the quantum Ising chain described by the Hamiltonian:
\begin{equation}
    H_{\text{Ising}}=-\sum_i Z_iZ_{i+1}-h_x\sum_iX_i-h_z\sum_i {Z_i},
\end{equation}
where $h_x$ and $h_z$ are the strength of the transverse and the longitudinal field. This quantun Ising chain can also be mapped to the $(1+1)$D $\mathbb{Z}_2$ lattice gauge theory~\cite{Gauging_Kitaev_2021}. The phase diagram of the quantum Ising chain is well-known, as shown in Fig.~\ref{Fig:Ising_chain}a. 

To detect the $Z$ 1-form symmetry, we first solve the ground states of the quantum Ising chain using the infinite MPS, from which we can create a periodic finite MPS with the perimeter $L$ using the tensor of the infinite MPS. We get the charge configurations by measuring the operators $\{Z_iZ_{i+1}\}$
from the finite periodic MPS. Then, we perform the simple QEC via the majority vote. An indictor of the $Z$ 1-form symmetry is simply the expectation value of a single $Z$ operator from the QEC recovered states.

We first consider the case of the transverse-field Ising model, i.e., $h_z=0$. The indictor is shown in Fig.~\ref{Fig:Ising_chain}c, implying that there is an emergent $Z$ 1-form symmetry in the ferromagnetic phase and there is no emergent $Z$ 1-form symmetry in the paramagnetic phase. This is consistent with the analytical results in Ref.~\cite{Sagar_ViJay_2024} and connects to the product-state limit discussed in the main text. 

We  also consider the case $h_z=0.2$. The indictor is shown in Fig.~\ref{Fig:Ising_chain}d, implying that there is always an emergent $Z$ 1-form symmetry when $h_z\neq 0$. Combining with the analytical analysis at the 1D product-states limit, we get the information-theoretic phase diagram of the $Z$ 1-form symmetry shown in Fig.~\ref{Fig:Ising_chain}b. On one hand, this example implies that our protocol is not limited to fine-tuned models. On the other hand, from a physical perspective it shows that in the $(1+1)$D $\mathbb{Z}_2$ lattice gauge theory, the confining regime is defined as the regime $h_z>0$, and the Higgs regime is defined as the regime $h_z=0$ and $h_x>1$ ($h_z=0$ and $h_x<1$ is the deconfined phase).

\section{Checking the modified detection protocol for planar geometries using various states}\label{sm:sec:modifiedind}
\begin{figure}[t]
     \centering
     \includegraphics[scale=0.5]{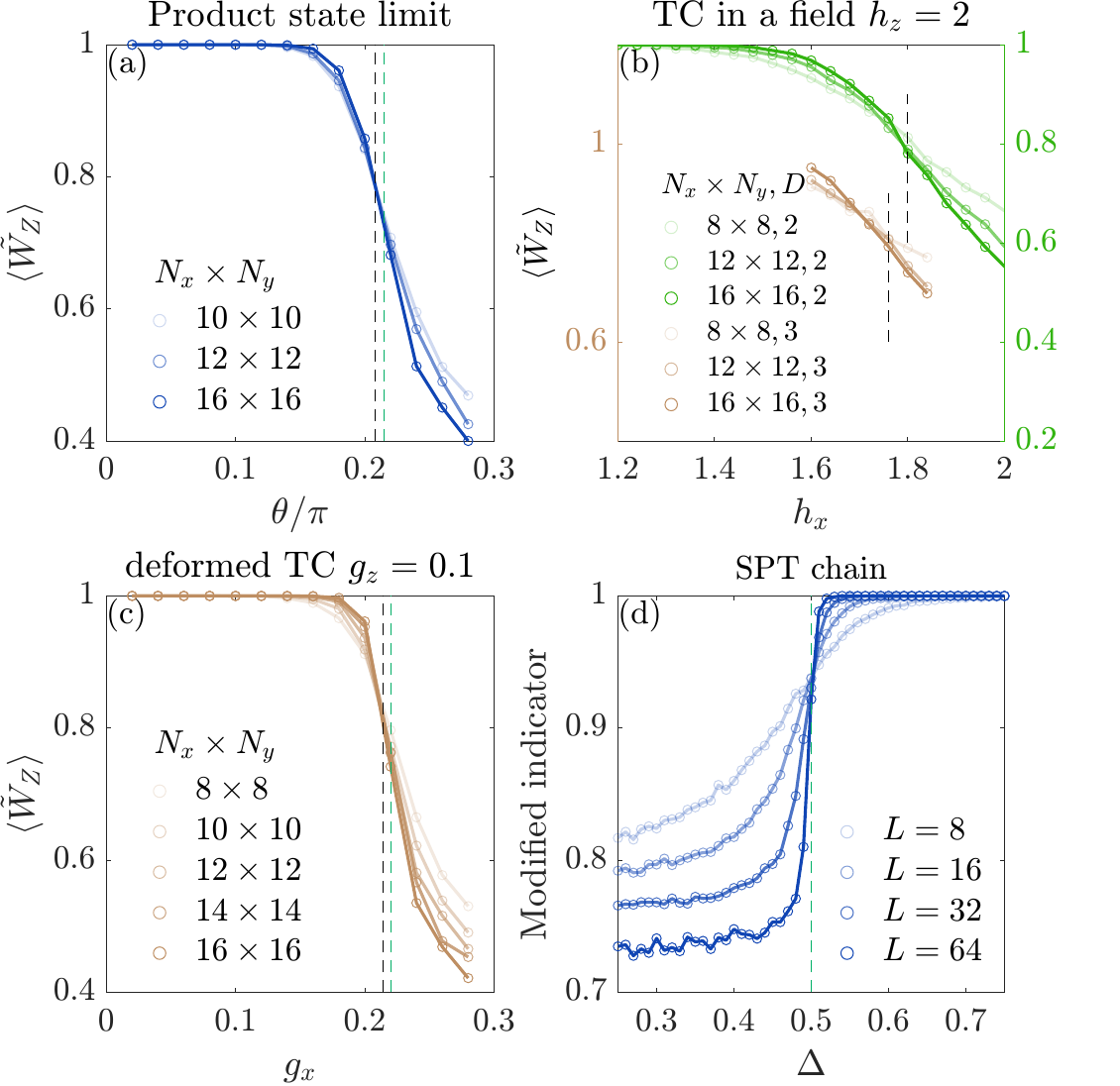}
     \caption{\textbf{Apply the modified indicator on a subsystem to various models.} 
     (a)  Indicator in the 2D product-state limit ($h_x^2+h_z^2=\infty$ or $g_x+g_z=1$) for various systems sizes. (b) Indicator along $h_z=2$ of the toric code model in a field in Eq.~\eqref{eq:TC_Hamiltonianwithfields} for various systems sizes and bond dimensions. The black dashed lines indicate the transition point from the unmodified indicator in Fig.~\ref{Fig:scan_main}d. (b) Indicator along $g_z=0.1$ of the deformed toric code model in Eq.~\eqref{eq:deformed_wavefunction} for various systems sizes. The black and the green dashed lines mark the numerical and the theoretical thresholds, respectively. (d) The modified indicator from the 1D $\mathbb{Z}_2\times\mathbb{Z}_2$ SPT states in Eq.~\eqref{eq:SPT_ham}. The results are obtained from a finite MPS with a length $2L$, and the subsystem size is $L$. The finite MPS is constructed from the infinite MPS tensor with a bond dimension $20$.}
     \label{Fig:open_system}
\end{figure}
In this section, to show that the modified indicator works well as expected, we apply the modified indicator to various 2D quantum states on the square lattice and to a 1D SPT model.

\textbf{2D quantum states.}
We consider an $N\times N$ subsystem $A$ embedded in the entire system on a cylinder with a size $2N\times N$ (see Fig.~\ref{Fig:protocol_for_subsystem}a), where $N=8,10,\cdots,16$. For numerical purposes the cylindrical geometry is beneficial over the previously discussed planar geometry. The basic ideas carry over. 
Figs.~\ref{Fig:open_system}a, b and c show the results of the modified indicator in the product-state limit $h_x^2+h_z^2=\infty$ ($g_x^2+g_z^2=1$), along the line $h_z=2$ or the toric code model in a field, and the line $g_z=0.1$ of the deformed toric code states, respectively. The modified indicators for different subsystem sizes approximately cross at a single point which agrees well with the threshold determined from the indicator evaluated on the torus geometry. A difference between the modified indicator and the previous one is that it does not tend to $0$ when the QEC fails, this problem can be fixed by choosing the size of the subsystem $A$ to be much smaller than the size of the entire system, instead of maintaining a constant ratio between the subsystem size and the entire system size.

\textbf{1D SPT states.}
The modified indicator introduced in Sec.~\ref{sec:protocol_for_open_sys} should in principle work for any problem with a 1-form symmetry or a similar structure. Besides the example of the deformed toric code, we show another example, which is a family of 1D $\mathbb{Z}_2\times\mathbb{Z}_2$ SPT states. We consider the 1D transverse-field cluster model describing a phase transition between a trivial $\mathbb{Z}_2\times\mathbb{Z}_2$ SPT phase and a non-trivial $\mathbb{Z}_2\times\mathbb{Z}_2$ SPT phase:
\begin{equation}\label{eq:SPT_ham}
    H=-(1-\Delta)\sum_iX_i-\Delta\sum_iZ_{i-1}X_iZ_{i+1}.
\end{equation}
The protecting $\mathbb{Z}_2\times\mathbb{Z}_2$ symmetry is generated by $\prod_iX_{2i}$ and  $\prod_iX_{2i+1}$. When $\Delta>0.5$ ($\Delta<0.5$), it is the non-trivial (trivial) SPT phase, and the phase transition point is at $\Delta=0.5$. 
Note that when only allowing the $\mathbb{Z}_2\times\mathbb{Z}_2$ symmetric perturbations, the cluster state (the ground state at $\Delta = 1$) shares a similar structure as the usual ferromagnetic Ising chain with the $Z$ 1-form symmetry, i.e. any symmetric perturbations to the ground state are generated by Pauli strings of $X$ and excite an even number of the cluster terms in Eq.~\eqref{eq:SPT_ham} located at the endpoints of the Pauli-$X$ string. The persistence of such a structure under any finite-depth symmetric local unitary perturbation has been previously exploited to probe the SPT phase transition using RG-type algorithms~\cite{RG_QEC_exact_2023}. Here, we detect the existence of the symmetry using the modified indicator based on a global QEC decoder. 

To examine the modified indicator for the 1D system, we approximate the ground state of the Hamiltonian in Eq.~\eqref{eq:SPT_ham} using the infinite MPS, from which we can create a periodic finite MPS using the tensor of the infinite MPS. We get the charge configurations by measuring the stabilizers $Z_{i-1}X_iZ_{i+1}$ on the finite MPS. Since the two $\mathbb{Z}_2$ symmetries apply on the even and odd sublattices separately, we separate the charges on the even and odd sublattices. Then, we do the first QEC by pairing the charges in the even (odd) sublattice separately. There are only two classes of inequivalent string operators for creating the excitations (i.e. the charges) in 1D for each sublattice, and we choose the one whose total length is the shortest.

We want to measure the modified indicator discussed in Sec.~\ref{sec:protocol_for_open_sys}. 
We proceed by performing the second QEC by considering a bi-partition of the system into two subsystems, $A$ and $B$. We remove those charges in $A$ connected to charges in $B$ by the strings from the first QEC. After that, we fold the subsystem $A$ into a ring and apply the second QEC to the folded subsystem $A$. We compare the output of the first QEC for charges in $A$ and the output of the second QEC and calculate the probability that the two QECs give the same string configuration connecting the remaining charges in $A$; see Fig.~\ref{Fig:open_system}d. The modified indicator increases (decreases) to $1$ (a number which is smaller than $1$) with the increasing of the size $L$ of the subsystem $A$ (the whole system size is $2L$), which correctly captures the phase transition of the transverse-field cluster model. To capture the correct criticality of the 1D SPT phase transition, we have to consider the case that the subsystem size is much smaller than the total system size. In Ref.~\cite{Sagar_ViJay_2024}, it is proposed that the detection of the 1D SPT phase transition can be achieved with a standard QEC requiring an open boundary condition with non-trivial ground-state degeneracy to encode quantum information. In contrast, our modified indicator can be applied to quantum states with any boundary condition.

\addtocontents{toc}{\protect\setcounter{tocdepth}{2}}
\bibliography{ref.bib}
\end{document}